\documentclass[a4paper]{article}
\pdfoutput=1
\usepackage{ifthen,relsize,multicol,marvosym,float}
\usepackage{tabularx,tabulary,multirow,lastpage,lineno}
\usepackage[pdftex]{graphicx}
\usepackage{epic,eepic,epsfig}
\usepackage{siunitx}
\usepackage[polutonikogreek,french,german,english]{babel}
\usepackage{chicago}

\usepackage{titling}
\usepackage[centertags]{amsmath}
\usepackage{pbox,fancyhdr,textcomp}
\usepackage{amssymb,amsfonts,oldgerm,mathrsfs,ipa}
\usepackage[text={6in,8.25in},centering]{geometry}

\newcommand{\hyperrefpdf}{\usepackage[colorlinks=true,citecolor=red,linkcolor=blue,urlcolor=blue,pdftex]{hyperref}}
\hyperrefpdf

\newcommand{\skipline}[1][1]{\vspace*{#1\baselineskip}}

\newcommand{\coloneq}{\mathrel{\mathop:}=}

\newcommand{\nth}[1][n]{$#1^{\text{th}}$}

\newcommand{\half}{\ensuremath{\frac{1}{2}}}

\newtheorem{theorem}{Theorem}[subsection]
\newtheorem{corollary}[theorem]{Corollary}
\newtheorem{lemma}[theorem]{Lemma}

\newtheorem{condition}[theorem]{Condition}
\newtheorem{defn}[theorem]{Definition}

\newtheorem{prop}[theorem]{Proposition}

\renewcommand{\thetheorem}{\arabic{section}.\arabic{theorem}}%
\renewcommand{\theequation}{\arabic{section}.\arabic{equation}}%
\renewcommand{\thefigure}{\arabic{section}.\arabic{figure}}%

\newcommand{\resetcounters}{%
  \setcounter{equation}{0}%
  \setcounter{theorem}{0}%
  \setcounter{figure}{0}%
}

\newcommand{\resetsec}{%
  \resetcounters%
  \renewcommand{\thetheorem}{\arabic{section}.\arabic{theorem}}%
  \renewcommand{\theequation}{\arabic{section}.\arabic{equation}}%
  \renewcommand{\thefigure}{\arabic{section}.\arabic{figure}}%
}

\newcommand{\french}[1]{\selectlanguage{french}#1\selectlanguage{english}}

\title{On Geometric Objects, the Non-Existence of a Gravitational
  Stress-Energy Tensor, and the Uniqueness of the Einstein Field
  Equation\thanks{This paper is published in \emph{Studies in History
      and Philosophy of Modern Physics}, 2019,
    doi:\href{http://dx.doi.org/10.1016/j.shpsb.2018.08.003}
    {10.1016/j.shpsb.2018.08.003}.}}

\author{Erik Curiel\thanks{I thank Robert Geroch for many stimulating
    conversations in which the seeds of several of the paper's ideas
    were germinated and, in some cases, fully cultivated to fruition.
    I also thank David Malament for helpful conversations on the
    principle of equivalence and on gravitational energy.  I am
    grateful to Ted Jacobson for commenting on an earlier draft and
    catching a serious error.  \textbf{Author's address}: Munich
    Center for Mathematical Philosophy,
    Ludwig-Maximilians-Universit\"at; Black Hole Initiative, Harvard
    University; Smithsonian Astrophysical Observatory, Radio and
    Geoastronomy Division; \textbf{email}:
    \href{mailto:erik@strangebeautiful.com}
    {\texttt{erik@strangebeautiful.com}}}}

\date{}

\begin{document}

\maketitle

\begin{quote}
  \begin{center}
    \textbf{ABSTRACT}
  \end{center}
  The question of the existence of gravitational stress-energy in
  general relativity has exercised investigators in the field since
  the inception of the theory.  Folklore has it that no adequate
  definition of a localized gravitational stress-energetic quantity
  can be given.  Most arguments to that effect invoke one version or
  another of the Principle of Equivalence.  I argue that not only are
  such arguments of necessity vague and hand-waving but, worse, are
  beside the point and do not address the heart of the issue.  Based
  on a novel analysis of what it may mean for one tensor to depend in
  the proper way on another, which, \emph{en passant}, provides a
  precise characterization of the idea of a ``geometric object'', I
  prove that, under certain natural conditions, there can be no tensor
  whose interpretation could be that it represents gravitational
  stress-energy in general relativity.  It follows that gravitational
  energy, such as it is in general relativity, is necessarily
  non-local.  Along the way, I prove a result of some interest in own
  right about the structure of the associated jet bundles of the
  bundle of Lorentz metrics over spacetime.  I conclude by showing
  that my results also imply that, under a few natural conditions, the
  Einstein field equation is the unique equation relating
  gravitational phenomena to spatiotemporal structure, and discuss how
  this relates to the non-localizability of gravitational
  stress-energy.  The main theorem proven underlying all the arguments
  is considerably stronger than the standard result in the literature
  used for the same purposes (Lovelock's theorem of 1972): it holds in
  all dimensions (not only in four); it does not require an assumption
  about the differential order of the desired concomitant of the
  metric; and it has a more natural physical interpretation.
\end{quote}

\skipline


\tableofcontents

\skipline[1.5]

\begin{quote}  
  As soon as the principle of conservation of energy was grasped, the
  physicist practically made it his definition of energy, so that
  energy was that something which obeyed the law of conservation.  He
  followed the practice of the pure mathematician, defining energy by
  the properties he wished it to have, instead of describing how he
  measured it.  This procedure has turned out to be rather unlucky in
  the light of the new developments.
  \begin{flushright}    
    Arthur Eddington \nocite{eddington-math-theor-rel} \\
    \emph{The Mathematical Theory of Relativity}, p.~136
  \end{flushright}
\end{quote}

\skipline

\section{Gravitational Energy in General Relativity}
\label{sec:intro}

There seems to be in general relativity no satisfactory, localized
representation of a quantity whose natural interpretation would be
``gravitational (stress-)energy''.  The only physically unquestionable
expressions of energetic quantities associated solely with the
``gravitational field'' we know of in general relativity are
quantities derived by integration over non-trivial volumes in
spacetimes satisfying any of a number of special
conditions.\footnote{\citeN[pp.~271--272]{weyl-space-time-matter} and
  \citeN[pp.~134--137]{eddington-math-theor-rel} were perhaps the
  first to grasp this point with real clarity.
  \citeN[pp.~104--105]{schrodinger-st-struc} gives a particularly
  clear, concise statement of the relation between the fact that the
  known energetic, gravitational quantities are non-tensorial and the
  fact that integration over them can be expected to yield integral
  conservation laws only under restricted conditions.}  These
quantities, moreover, tend to be non-tensorial in character.  In other
words, these are strictly non-local quantities, in the precise sense
that they are not represented by invariant geometric objects defined
at individual spacetime points (such as tensors or scalars).

This puzzle about the character and status of gravitational energy
emerged simultaneously with the discovery of the theory
itself.\footnote{The first pseudo-tensorial entity proposed to
  represent gravitational stress-energy dates back to
  \citeN{einstein-gr}, the paper in which he first proposed the final
  form of the theory.}  The problems raised by the seeming
non-localizability of gravitational energy had a profound, immediate
effect on subsequent research.  It was, for instance, directly
responsible for Hilbert's request to Noether that she investigate
conservation laws in a quite general setting, the work that led to her
famous results relating symmetries and conservation laws
\cite{brading-energy-cons-gr}.

Almost all discussions of gravitational energy in general relativity,
however, dating back even to the earliest ones, have been plagued by
vagueness and lack of precision.  The main result of this paper
addresses the issue head-on in a precise and rigorous way.  Based on
an analysis of what it may mean for one tensor to depend in the proper
way on another, I prove that, under certain natural conditions, there
can be no tensor whose interpretation could be that it represents
gravitational stress-energy in general relativity.  It follows that
gravitational stress-energy, such as it is in general relativity, is
necessarily non-local.  Along the way, I prove a result of some
interest in its own right about the structure of the associated first
two jet bundles of the bundle of Lorentz metrics over spacetime.  I
conclude with a discussion of the sense in which my results also show
that the Einstein field equation is, in a natural sense, the unique
field equation in the context of a theory such as general relativity,
and discuss how this fact relates to the non-localizability of
gravitational stress-energy.

The main theorem (\ref{thm:only-einstein-tens}) underlying all the
arguments is considerably stronger than the standard result in the
literature used to argue for the uniqueness of the Einstein field
equation (the classic theorem of \citeNP{lovelock-4dim-space-efe},
stated in footnote~\ref{fn:lovelock}): it holds in all dimensions, not
only in four; it does not require an assumption about the differential
order of the desired concomitant of the metric; and it has a more
natural physical interpretation.  The theorem also has interesting
consequences for a proper understanding of the cosmological-constant
term in the Einstein field equation, and for higher-dimensional
Lanczos-Lovelock theories of gravity, which I discuss at the end of
the paper.

\section{The Principle of Equivalence:  A Bad Argument}
\label{sec:princ_equiv}
\resetsec

The most popular heuristic argument used to attempt to show that
gravitational energy either does not exist at all or does exist but
cannot be localized invokes the ``principle of equivalence''.
\citeN[p.~399]{choquet83}, for example, puts the argument like this:
\begin{quote}
  This `non local' character of gravitational energy is in fact
  obvious from a formulation of the equivalence principle which says
  that the gravitational field appears as non existent to one observer
  in free fall.  It is, mathematically, a consequence of the fact that
  the pseudo-riemannian connexion which represents the gravitational
  field can always be made to vanish along a given curve by a change
  of coordinates.
\end{quote}
\citeN[pp.~135-6]{trautman-energy-grav-cosmo} and
\citeN[pp.~469-70]{goldberg-inv-trans-cons-laws-ener-mom} also made
essentially the same
argument.\footnote{\label{fn:fungible-se-tensor}Goldberg's formulation
  of the argument makes explicit a feature at least implicitly common
  in the many instances I have found in the literature, the conclusion
  that a local gravitational energy \emph{scalar} density does not
  exist and not that a gravitational stress-energy tensor does not
  exist.  One cannot have a scalar energy density for a physical field
  in general relativity, however, without an associated stress-energy
  tensor.  Such a state of affairs would violate the thermodynamic
  principle that all energy is equivalent in character, in the sense
  that any one form can always in principle be tranformed into any
  other form, since all other known forms of physical field do have a
  stress-energy tensor as the fundamental representation of their
  energetic content.  I discuss this in more detail in
  \S\ref{sec:conds}, especially footnote~\ref{fn:other-energy-qs}.}
Indeed, the making of this argument seems to be something of a shared
mannerism among physicists who discuss gravitational energy in general
relativity; it is difficult to find an article on the topic in which
it is not at least alluded
to.\footnote{\citeN{bondi-phys-char-grav-wvs},
  \citeN{penrose-gr-enflux-elem-opt} and \citeN{geroch-energy-extrac}
  are notable exceptions.  I take their discussions as models of how
  one should discuss energetic phenomena in the presence of
  gravitational fields.}

The argument has a fundamental flaw.  It assumes that, if there is
such a thing as localized gravitational energy or stress-energy, it
can depend only on ``first derivatives of the metric''---that those
first derivatives encode all information about the ``gravitational
field'' relevant to stress-energy---for it is only entities depending
only on those first derivatives that one can make vanish along curves.
But that seems wrong on the face of it.  If there is such a thing as a
localized gravitational energetic quantity, then surely it depends on
the curvature of spacetime and not on the affine connection (or, more
precisely, it depends on the affine connection at least in so far as
it depends on the curvature), for any energy one can envision
transferring from the gravitational field to another type of system in
a different form in general relativity (\emph{e}.\emph{g}., as heat or
a spray of fundamental particles) must at bottom be based on geodesic
deviation,\footnote{\citeN{penrose-gr-enflux-elem-opt} and
  \citeN{ashtekar-penrose-mass-pos-focus-struc-sl-inf} rely on the
  same idea to very fruitful effect.} and so must be determined by the
value of the Riemann tensor at a point, not by the value of the affine
connection at a point or even along a curve.  There is no solution to
the Einstein field-equation that corresponds in any natural way to the
intuitive Newtonian idea of a constant non-zero gravitational field,
\emph{i}.\emph{e}., one without geodesic deviation; that, however,
would be the only sort of field that one could envision even being
tempted to ascribe gravitational energy to in the absence of geodesic
deviation, and that attribution is problematic even in Newtonian
theory.  Indeed, a spacetime has no geodesic deviation if and only if
it is everywhere locally isometric to Minkowski spacetime, which we
surely want to say has vanishing gravitational energy if any spacetime
does, if one can make such a statement precise in the first
place.\footnote{One might be tempted by the stronger claim that
  Minkowski spacetime ought to be the unique spacetime with vanishing
  gravitational energy.  I do not think that can be right, however.
  If the existence of gravitational energy is indeed intimately tied
  with the presence of geodesic deviation (as argued forcefully by
  \citeNP{penrose-gr-enflux-elem-opt}), then any flat spacetime, such
  as that of \citeN{kasner-geom-thms-efe}, also ought to have
  vanishing gravitational energy.}

An obvious criticism of my response to the standard line, related to a
popular refinement of the argument given for the non-existence or
non-locality of gravitational energetic quantities, is that it would
make gravitational stress-energy depend on second-order partial
derivatives of the field potential (the metric, so comprehended by
analogy with the potential in Newtonian theory), whereas all other
known forms of stress-energy depend only on terms quadratic in the
first partial derivatives of the field potential.  To be more precise,
the argument runs like this:
\begin{quote}
  One can make precise the sense in which Newtonian gravitational
  theory is the ``weak-field'' limit of general relativity
  \cite{malament-newt-grav-geom-spc}.  In this limit, it is clear that
  the metric field plays roughly the role in general relativity that
  the scalar potential $\phi$ does in Newtonian theory.  In Newtonian
  theory, bracketing certain technical questions about boundary
  conditions, there is a more or less well-defined energy density of
  the gravitational field, proportional to $( \nabla \phi )^2$.  One
  might expect, therefore, based on some sort of continuity argument,
  or just on the strength of the analogy itself, that any local
  representation of gravitational energy in general relativity ought
  to be a ``quadratic function of the first partials of the
  metric''.\footnote{In this light, it is interesting to note that
    gravitational energy pseudo-tensors do tend to be quadratic in the
    first-order partials of the metric
    \cite{einstein-gr,moller-thry-rel-2nd,landau-lifschitz-fields}.}
  The stress-energy tensor of no other field, moreover, is higher than
  first-order in the partials of the field potential, so surely
  gravity cannot be different.  No invariant quantity at a point can
  be constructed using only the first partials of the metric, however,
  so there can be no scalar or tensorial representation of
  gravitational energy in general relativity.
\end{quote}
(No researcher I know makes the argument exactly in this form; it is
just the clearest, most concise version I can come up with myself.)
As \citeN[p.\ 178]{pauli-thry-rel} forcefully argued, however, there can be
no \emph{physical} argument against the possibility that gravitational
energy depends on second derivatives of the metric; the argument above
certainly provides none.  Just because the energy of all other known
fields have the same form in no way implies that a localized
gravitational energy in general relativity, if there is such a thing,
ought to have that form as well.  Gravity is too different a field
from others for such a bare assertion to carry any weight.  As I
explain at the end of \S\ref{sec:conds}, moreover, a proper
understanding of tensorial concomitants reveals that an expression
linear in second partial derivatives is in the event equivalent in the
relevant sense to one quadratic in first order partials.  This
illustrates how misleading the analogy with Newtonian gravity can be.

\section{Geometric Fiber Bundles, Concomitants, and Geometric
  Objects}
\label{sec:concomitants}
\resetsec

\skipline[.5]

\begin{quote}
  The introduction of a coordinate system to geometry is an act of
  violence.
  \skipline[-.5]
  \begin{flushright}
    Hermann Weyl \\
    \emph{Philosophy of Mathematics and Natural Science}
  \end{flushright}
\end{quote}

I have argued that, if there is an object that deserves to be thought
of as the representation of gravitational stress-energy in general
relativity, then it ought to depend on the Riemann curvature tensor.
Since there is no obvious mathematical sense in which a general
mathematical structure can ``depend'' on a tensor, the first task is
to say what exactly this could mean.  I will call a mathematical
structure on a manifold that depends in the appropriate fashion on
another structure on the manifold, or set of others, a
\emph{concomitant} of it (or them).

The reason I am inquiring into the possibility of a concomitant in the
first place, when the question is the possible existence of a
representation of gravitational stress-energy tensor, is a simple one.
What is wanted is an expression for gravitational energy that does not
depend for its formulation on the particulars of the spacetime, just
as the expression for the kinetic energy of a particle in classical
physics does not depend on the internal constitution of the particle
or on the particular interactions it may have with its environment,
and just as the stress-energy tensor for a Maxwell field has the same
form as a function of the Faraday tensor in every spacetime
irrespective of its particulars.\footnote{This property of
  (stress-)energy for other types of physical systems already stands
  in contradistinction to the properties of all known rigorous
  expressions for global gravitational energy in general relativity,
  \emph{e}.\emph{g}., the ADM mass and the Bondi energy, which can be
  defined only in asymptotically flat spacetimes \cite{wald-gr}, and
  all such quasi-local expressions, which can be defined only in
  stationary or axisymmetric ones
  \cite{szabados-qloc-en-mom-ang-mom-gr}.}  If there is a well-formed
expression for gravitational stress-energy, then one should be able in
principle to calculate it whenever there are gravitational phenomena,
which is to say, in any spacetime whatsoever---it should be a
\emph{function} of some set of geometric objects associated with the
curvature in that spacetime, in some appropriately generalized sense
of `function'.  This idea is what a concomitant is supposed to
capture.

The term `concomitant' and the general idea of the thing is due to
\citeN[p.~15]{schouten-ricci-calc}.\footnote{The specific idea of
  proving the uniqueness of a tensor that ``depends'' on another
  tensor, and satisfies a few collateral conditions, dates back at
  least to \citeN[pp.~315--318]{weyl-space-time-matter} and
  \citeN{cartan22}.  In fact, Weyl proved that the only two-index
  symmetric covariant tensors one can construct at a point in any
  spacetime, using only algebraic combinations of the components of
  the metric and its first two partial derivatives in a coordinate
  system at that point, that are at most linear in the second
  derivatives of the metric, are linear combinations of the Ricci
  curvature tensor, the scalar curvature times the metric and the
  metric itself.  In particular, the only such divergence-free tensors
  one can construct at a point are linear combinations of the Einstein
  tensor and the metric with constant coefficients.}  The definition
Schouten proposed is expressed in terms of coordinates: depending on
what sort of concomitant one deals with, the components of the
concomitant in a given coordinate system must satisfy various
conditions of covariance under certain classes of coordinate
transformations, when those transformations are also applied to the
components of the objects the concomitant is defined as a ``function''
of.  His work was picked up and generalized by several other
mathematicians, such as \citeN{aczel-komit-diff-komit}, who extended
Schouten's work to treat more generalized classes of higher-order
differential concomitants.\footnote{I thank an anonymous referee for
  drawing my attention to the work of Acz\'el and others who developed
  Schouten's work.}  \label{pg:schouten-difficult}The definitions
provided by this early work is clear, straightforward and easy to
grasp in the abstract, but becomes difficult to work with in
particular cases of interest---Schouten's covariance conditions
translate into a set of partial differential equations in a particular
coordinate system, which even in seemingly straightforward cases turn
out to be forbiddingly complicated.  This makes it not only unwieldy
in practice and inelegant, but, more important, it makes it difficult
to discern what of intrinsic physical significance is encoded in the
relation of being a concomitant in particular cases of interest.  It
is almost impossible to determine anything of the general properties
of a particular kind of concomitant of a particular (set of) object(s)
by looking at those equations.\footnote{\label{fn:hairy}For a good
  example of just how hairy those conditions can be, see
  \citeN[p.~350]{duplessis-tens-concoms-cons-laws} for a complete set
  written out explicitly in the case of two covariant-index tensorial
  second-order differential concomitants of a Lorentz metric.}  I
suspect that it is because in particular cases the conditions are so
complex, difficult and opaque that use is very rarely made of
concomitants in arguments about spacetime structure in general
relativity.  This is a shame, for the idea is, I think, potentially
rich, and so calls out for an invariant formulation.\footnote{There is
  a tradition, initiated in the 1970s by \citeN{nijenhuis-natl-bunds},
  that attempts a more invariant formulation of a notion similar to
  Schouten's original one, introducing the idea of ``natural bundles''
  as a setting for the definition and study of structures closely
  related to what I call here geometrical objects.  That work was
  elaborated and extended by, \emph{e}.\emph{g}.,
  \citeN{epstein-natl-tens-riem-mnflds} and
  \citeN{epstein-thurston-trans-grps-natl-bunds}, \emph{inter alia}.
  That work is similar to the constructions and arguments I give here.
  I did not know of it when I developed my own work.  (Again, I thank
  the anonymous referee for drawing my attention to it.)  There are
  two novelties I can claim for my definitions and constructions
  (besides the fact that it is now all presented in a purely invariant
  way, with no use of coordinates).  First is my definition of fiber
  bundles without reference to an associated group of transformations,
  and so the consequent development of what I call geometric bundles
  based on the idea of inductions.  Second, the idea of an induction
  allows for a simple generalization of my definition for concomitants
  to more general structures than just tensorial-like objects,
  \emph{e}.\emph{g}., projective structures as characterized by an
  appropriate family of curves; I do not develop that generalization
  here as it is not needed.  Also, to the best of my knowledge, the
  main result of \S\ref{sec:concomitants-metric},
  theorem~\ref{thm:1-2-jet-metric}, is new, and of some interest in
  its own right, besides the use I put it to in proving
  theorem~\ref{thm:only-einstein-tens}.  (There is some contemporary
  work being done on so-called natural
  transformations---\emph{e}.\emph{g}., \citeNP{kolar-et-nat-opns-dg}
  and \citeNP{fatibene-francaviglia-nat-gauge-form-cft}, dating back
  to \citeNP{palais-terng-nat-bunds-fin-ord}---that bears some
  similarity to all these ideas, but I do not discuss it, first
  because it is formulated in category theory and so is fundamentally
  algebraic in nature, whereas I aim for a formulation with clear and
  intuitive geometric content, and second because my idea of an
  induction differentiates my work in important ways from it.)}

I use the machinery of fiber bundles to characterize the idea of a
concomitant in invariant terms.  I give a (brief) explicit formulation
of the machinery, because the one I rely on is non-standard.  (We
assume from hereon that all relevant structures, mappings,
\emph{etc}., are smooth; nothing is lost by the assumption and it
simplifies exposition---all germane constructions and proofs can
easily be generalized to the case of topological spaces and continuous
structures.)
\begin{defn}
  \label{defn:fiberbundle}
  A \emph{fiber bundle} $\mathfrak{B}$ is an ordered triplet,
  $(\mathcal{B}, \, \mathcal{M}, \, \pi)$, such that:
  \begin{enumerate}
    \addtolength{\itemindent}{2em} 
      \item[\bf{FB1}.] $\mathcal{B}$ is a differential manifold
      \item[\bf{FB2}.] $\mathcal{M}$ is a differential manifold
      \item[\bf{FB3}.] $\pi : \mathcal{B} \rightarrow \mathcal{M}$ is
    smooth and onto
      \item[\bf{FB4}.] For every $q,p \in \mathcal{M}$, $\pi^{-1} ( q
    )$ is diffeomorphic to $\pi^{-1} ( p )$ (as submanifolds of
    $\mathcal{B}$)
      \item[\bf{FB5}.] $\mathcal{B}$ has a locally trivial product
    structure, in the sense that for each $q \in \mathcal{M}$ there is
    a neighborhood $U \ni q$ and a diffeomorphism $\zeta : \pi^{-1}
    [U] \rightarrow U \times \pi^{-1} (q)$ such that the action of
    $\pi$ commutes with the action of $\zeta$ followed by projection
    on the first factor.
  \end{enumerate}
\end{defn}
$\mathcal{B}$ is the \emph{bundle space}, $\mathcal{M}$ the \emph{base
  space}, $\pi$ the \emph{projection} and $\pi^{-1} ( q )$ the
\emph{fiber} over $q$.  By a convenient, conventional abuse of
terminology, I will sometimes call $\mathcal{B}$ itself `the fiber
bundle' (or `the bundle' for short).  A \emph{cross-section} $\kappa$
is a smooth map from $\mathcal{M}$ into $\mathcal{B}$ such that $\pi
(\kappa (q)) = q$, for all $q$ in the mapping's domain.

This definition of a fiber bundle is non-standard in so far as no
group action on the fibers is fixed from the start; this implies that
no correlation between diffeomorphisms of the base space and
diffeomorphisms of the bundle space is fixed.\footnote{See,
  \emph{e}.\emph{g}., \citeN{steenrod-topo-fbs} for the traditional
  definition and the way that a fixed group action on the fibers
  induces a correlation between diffeomorphisms on the bundle space
  and those on the base space.}  One must fix that explicitly.  On the
view I advocate, the geometric character of the objects represented by
the bundle arises arises not from the group action directly, but only
after the explicit fixation of a correlation between diffeomorphisms
on the base space with those on the bundle space---only after, that
is, one fixes how a diffeomorphism on the base space induces one on
the bundle.  For example, depending on how one decides that a
diffeomorphism on the base space ought to induce a diffeomorphism on
the bundle over it whose fibers consist of 1-dimensional vector
spaces, one will ascribe to the objects of the bundle the character
either of ordinary scalars or of $n$-forms (where $n$ is the dimension
of the base space).  The idea is that the diffeomorphisms induced on
the bundle space then implicitly define the group action on the fibers
appropriate for the required sort of object.\footnote{I will not work
  out here the details of how this comes about, as they are not needed
  for the arguments of the paper.}

I call an appropriate mapping of diffeomorphisms on the base space to
those on the bundle space an \emph{induction}.  (I give a precise
definition in a moment.)  In this scheme, therefore, the induction
comes first conceptually, and the relation between diffeomorphisms on
the base space and those they induce on the bundle serves to fix the
fibers as spaces of \emph{geometric objects}, \emph{viz}., those whose
transformative properties are tied directly and intimately to those of
the ambient base space.\footnote{See
  Anderson~\citeyear{anderson62,anderson-princs-rel},
  \citeN{friedman-fnds-st-theors} and \citeN{belot-bckgrnd-indep} for
  other approaches to defining geometric or (as they refer to them)
  absolute objects.}  This way of thinking of fiber bundles is perhaps
not well suited to the traditional mathematical task of classifying
bundles, but it turns out to be just the thing on which to base a
perspicuous and useful definition of concomitant.  Although a
diffeomorphism on a base space will naturally induce a unique one on
certain types of fiber bundles over it, such as tensor bundles, in
general it will not.  There is not known, for instance, any natural
way to single out a map of diffeomorphisms of the base space into
those of a bundle over it whose fibers consist of spinorial
objects.\footnote{See, \emph{e}.\emph{g}.,
  \citeN{penrose-rindler-spinors-st-1}.}  Inductions neatly handle
such problematic cases.

I turn now to making this intuitive discussion more precise.  A
diffeomorphism $\phi^\sharp$ of a bundle space $\mathcal{B}$ is
\emph{consistent} with $\phi$, a diffeomorphism of the base space
$\mathcal{M}$, if, for all $u \in \mathcal{B}$,
\[
\pi (\phi^\sharp (u)) = \phi (\pi (u))
\] 
For a general bundle, there will be scads of diffeomorphisms
consistent with a given diffeomorphism on the base space.  A way is
needed to fix a unique $\phi^\sharp$ consistent with a $\phi$ so that
a few obvious conditions are met.  For example, the identity
diffeomorphism on $\mathcal{M}$ ought to pick out the identity
diffeomorphism on $\mathcal{B}$.  More generally, if $\phi$ is a
diffeomorphism on $\mathcal{M}$ that is the identity on an open set $O
\subset \mathcal{M}$ and differs from the identity outside $O$, it
ought to be the case that the mapping picks out a $\phi^\sharp$ that
is the identity on $\pi^{-1} [O]$.  If this holds, we say that that
$\phi^\sharp$ is \emph{strongly consistent} with $\phi$.

Let $\mathfrak{D}_\mathcal{M}$ and $\mathfrak{D}_\mathcal{B}$ be,
respectively, the groups of diffeomorphisms on $\mathcal{M}$ and
$\mathcal{B}$.  Define the set
\begin{center}
  $\mathfrak{D}^\sharp_\mathcal{B} = \{ \phi^\sharp \in
  \mathfrak{D}_\mathcal{B} : \: \exists \phi \in
  \mathfrak{D}_\mathcal{M}$ such that $\phi^\sharp$ is strongly
  consistent with $\phi \}$
\end{center}
It is simple to show that $\mathfrak{D}^\sharp_\mathcal{B}$ forms a
subgroup of $\mathfrak{D}_\mathcal{B}$.  This suggests 
\begin{defn}
  \label{defn:induction}
  An \emph{induction} is an injective homomorphism $\iota :
  \mathfrak{D}_\mathcal{M} \rightarrow
  \mathfrak{D}^\sharp_\mathcal{B}$.
\end{defn}
$\phi$ will be said to \emph{induce} $\phi^\sharp$ (under $\iota$) if
$\iota ( \phi ) = \phi^\sharp$.\footnote{In a more thorough treatment,
  one would characterize the way that the induction fixes a group
  action on the fibers, but we do not need to go into that for our
  purposes.}
\begin{defn}
  \label{defn:geom-bundle}
  A \emph{geometric fiber bundle} is an ordered quadruplet
  $(\mathcal{B}, \, \mathcal{M}, \, \pi, \, \iota)$ where
  \begin{enumerate}
    \addtolength{\itemindent}{3em} 
      \item[{\bf GFB1}.] $(\mathcal{B}, \, \mathcal{M}, \, \pi)$
    satisfies FB1-FB5
      \item[{\bf GFB2}.] $\iota$ is an induction
  \end{enumerate}
\end{defn} 
Geometric fiber bundles are the appropriate spaces to serve as the
domains and ranges of concomitant mappings.

Most of the fiber bundles one works with in physics are geometric
fiber bundles.  A tensor bundle $\mathcal{B}$, for example, is a fiber
bundle over a manifold $\mathcal{M}$ each of whose fibers is
diffeomorphic to the vector space of tensors of a particular index
structure over any point of the manifold; a basis for an atlas is
provided by the charts on $\mathcal{B}$ naturally induced from those
on $\mathcal{M}$ by the representation of tensors on $\mathcal{M}$ as
collections of components in $\mathcal{M}$'s coordinate systems.
There is a natural induction in this case fixed by the pull-back
action of a diffeomorphism $\phi$ of tensors on $\mathcal{M}$.  Spinor
bundles provide interesting examples of physically important bundles
that have no natural, unique inductions, though there are classes of
them.

We are finally in a position to define concomitants.  Let
$(\mathcal{B}_1, \, \mathcal{M}, \, \pi_1, \, \iota_1)$ and
$(\mathcal{B}_2, \, \mathcal{M}, \, \pi_2, \, \iota_2)$ be two
geometric bundles with the same base space.\footnote{One can
  generalize the definition of concomitants to cover the case of
  bundles over different base spaces, but we do not need this here.}
\begin{defn}
  \label{def:concom}
  A mapping $\chi : \mathcal{B}_1 \rightarrow \mathcal{B}_2$ is a
  \emph{concomitant} if
  \[
    \chi (\iota_1 (\phi) (u_1)) = \iota_2 (\phi) (\chi (u_1))
  \]
  for all $u_1 \in \mathcal{B}_1$ and all $\phi \in
  \mathfrak{D}_\mathcal{M}$.
\end{defn}
In intuitive terms, a concomitant is a mapping between bundles that
commutes with the action of the induced diffeomorphisms that lend the
objects of the bundles their respective geometrical characters,
\emph{i}.\emph{e}., the structure in virtue of which they are, in a
precise sense, \emph{geometric} objects.  It is easy to see that
$\chi$ must be fiber-preserving, in the sense that it maps fibers of
$\mathcal{B}_1$ to fibers of $\mathcal{B}_2$.  This captures the idea
that the dependence of the one type of object on the other is strictly
local; the respecting of the actions of diffeomorphisms captures the
idea that the mapping encodes an invariant relation.  By another
convenient abuse of terminology, I will often refer to the range of
the concomitant mapping itself as `the concomitant' of the domain.

\section{Jet Bundles, Higher-Order Concomitants, and Geometric
  Objects}
\label{sec:jet-bundles}
\resetsec

Just as with ordinary functions from one Euclidean space to another,
it seems plausible that the dependence encoded in a concomitant from
one geometric bundle to another may take into account not only the
value of the first geometrical structure at a point of the base space,
but also ``how that value is changing'' in a neighborhood of that
point, something like a generalized derivative of a geometrical
structure on a manifold.  The following construction is meant to
capture in a precise sense the idea of a generalized derivative in
such a way so as to make it easy to generalize the idea of a
concomitant to account for it.

Fix a geometric fibre bundle
$(\mathcal{B}, \, \mathcal{M}, \, \pi, \, \iota)$, and the space of
its sections $\Gamma [\mathcal{B}]$.  Two sections
$\gamma, \eta : \mathcal{M} \rightarrow \mathcal{B}$ \emph{osculate to
  first-order} at $p \in \mathcal{M}$ if $T\gamma$ and $T\eta$ (the
differentials of the mappings) agree in their action on
$T_p \mathcal{M}$.\footnote{See, \emph{e}.\emph{g}.,
  \citeN[p.~18]{hirsch-diff-top} for the definition of the
  differential of a mapping.}  (They osculate to zeroth-order at $p$
if they map $p$ to the same point in the domain.)  This defines an
equivalence relation on $\Gamma [\mathcal{B}]$.  A \emph{1-jet} with
source $p$ and target $\gamma(p)$, written `$j^1_p [\gamma]$', is such
an equivalence class.  It is not difficult to show that the set of all
1-jets,
\[
J^1 \mathcal{B} \coloneq \bigcup_{p \in \mathcal{M}, \gamma \in \Gamma
  [\mathcal{B}]} j^1_p [\gamma]
\]
naturally inherits the structure of a differentiable manifold
\cite{hirsch-diff-top}.  One can naturally fibre $J^1 \mathcal{B}$
over $\mathcal{M}$.  The \emph{source projection}
$\sigma^1 : J^1 \mathcal{B} \rightarrow \mathcal{M}$, defined by
\[
\sigma^1 (j^1_p [\gamma]) = p
\]
gives $J^1 \mathcal{B}$ the structure of a bundle space over the base
space $\mathcal{M}$, and in this case we write the bundle $(J^1
\mathcal{B}, \, \mathcal{M}, \, \sigma^1)$.  A section $\gamma$ of
$\mathcal{B}$ naturally gives rise to a section $j^1 [\gamma]$ of $J^1
\mathcal{B}$, the \emph{first-order prolongation} of that section:
\[
j^1 [\gamma] : \mathcal{M} \rightarrow \bigcup_{p \in \mathcal{M}}
j^1_p [\gamma]
\]
such that $\sigma_1 (j^1 [\gamma] (p)) = p$.  (We assume for the sake
of simplicity that global cross-sections exist; the modifications
required to treat local cross-sections are trivial, albeit tedious.)

The points of $J^1 \mathcal{B}$ may be thought of as coordinate-free
representations of first-order Taylor expansions of sections of
$\mathcal{B}$.  To see this, consider the example of the trivial
bundle $(\mathcal{B}, \, \mathbb{R}^2, \, \pi)$ where $\mathcal{B}
\coloneq \mathbb{R}^2 \times \mathbb{R}$ and $\pi$ is projection onto
the first factor.  Fix global coordinates $(x^1, \, x^2, \, v^1)$ on
$\mathcal{B}$, so that the induced (global) coordinates on $J^1
\mathcal{B}$ are $(x^1, \, x^2, \, v^1, \, v^1_1, \, v^1_2)$.  Then
for any 1-jet $j^1_q [\gamma]$, define the inhomogenous linear
function $\hat{\gamma} : \mathbb{R}^2 \rightarrow \mathbb{R}$ by
\[
\hat{\gamma}(p) = v^1 (\gamma(p)) + v^1_1 (j^1_q [\gamma])(p_1 - q_1)
+ v^1_2 (j^1_q [\gamma])(p_2 - q_2)
\]
where $\gamma \in j^1_q [\gamma]$, and $p, q \in \mathbb{R}^2$ with
respective components $(p_1, \, p_2)$ and $(q_1, \, q_2)$.  Clearly
$\hat{\gamma}$ defines a cross-section of $J^1 \mathcal{B}$
first-order osculant to $\gamma$ at $p$ and so is a member of $j^1_q
[\gamma]$; indeed, it is the unique globally defined, linear
inhomogeneous map with this property.

A 2-jet is defined similarly, by iteration, as an equivalence class of
sections under the relation of having the same first and second
differentials (as mappings) at a point.  More precisely,
$\gamma, \eta \in \Gamma [\mathcal{B}]$ \emph{osculate to second
  order} at $p \in \mathcal{M}$ if they are in the same 1-jet and if
their second-order differentials equal each other,
$T(T\gamma) = T(T\eta)$.  Again, this defines an equivalence relation
on $\Gamma [\mathcal{B}]$.  A \emph{2-jet} with source $p$ and target
$\gamma(p)$, written `$j^1_p [\gamma]$', is such an equivalence class.
The set of all 2-jets,
\[
J^2 \mathcal{B} \coloneq \bigcup_{p \in \mathcal{M}, \gamma \in \Gamma
  [\mathcal{B}]} j^2_p [\gamma]
\]
also inherits the structure of a differentiable manifold.
$J^2 \mathcal{B}$ is naturally fibered over $\mathcal{M}$ by the
source projection
$\sigma^2 : J^2 \mathcal{B} \rightarrow \mathcal{M}$, defined by
\[
\sigma^2 (j^2_p [\gamma]) = p
\]
giving $J^2 \mathcal{B}$ the structure of a bundle space over the base
space $\mathcal{M}$, $(J^2 \mathcal{B}, \, \mathcal{M}, \, \sigma^2)$.
Again, a section $\gamma$ of $\mathcal{B}$ gives rise to a section
$j^2 [\gamma]$ of $J^2 \mathcal{B}$, the \emph{second-order
  prolongation} of that section:
\[
j^2 [\gamma] : \mathcal{M} \rightarrow \bigcup_{p \in \mathcal{M}}
j^2_p [\gamma]
\]
such that $\sigma_1 (j^2 [\gamma] (p)) = p$.  Jet bundles of all
higher orders are defined in the same way.

There is a natural projection from $J^2 \mathcal{B}$ to
$J^1 \mathcal{B}$, the \emph{truncation} $\theta^{2,1}$, characterized
by ``dropping the second-order terms in the Taylor expansion''.  More
precisely, for $j^2 [\gamma]$, the truncation is the unique
$j^1 [\eta]$ such that $T j^1 [\eta] = T T \gamma$, which guarantees
that $j^1 [\eta] = j^1 [\gamma]$.\footnote{One might worry that the
  truncation is not unique, because two 1-jets may ``differ only by a
  constant'' and so still give the same 2-jet, as may happen with
  ordinary derivatives in calculus.  Because there is no privileged
  derivative operator on $J^1 \mathcal{B}$, however, there is no well
  defined notion of two 1-jets ``differing by a constant''.}  In
general, one has the natural truncation
$\theta^{n,m} : J^n \mathcal{B} \rightarrow J^m \mathcal{B}$ for all
$0 < m < n$.

For our purposes, the most important fact about these spaces is that
the jet bundles of a geometric bundle are themselves naturally
geometric bundles.  Fix a geometric bundle
$(\mathcal{B}, \, \mathcal{M}, \pi, \, \iota)$ and a diffeomorphism
$\phi$ on $\mathcal{M}$.  Then $\iota[\phi]$ not only defines an
action on points of $\mathcal{B}$, but, as a diffeomorphism itself on
$\mathcal{B}$, it naturally defines an action on the cross-sections of
$\mathcal{B}$ and thus on the 1-jets. by the natural pull-back of
differentials of mappings.  It is easy to show that the mapping
$\iota^1$ so specified from $\mathfrak{D}_\mathcal{M}$ to
$\mathfrak{D}^\sharp_{J^1 \mathcal{B}}$ is an injective homomorphism
and thus itself an induction; therefore,
$(J^1 \mathcal{B}, \, \mathcal{M}, \sigma^1, \, \iota^1)$ is a
geometric fiber bundle.  One defines inductions for higher-order jet
bundles in the same way.

We can now generalize our definition of concomitants.  Let
$(\mathcal{B}_1, \, \mathcal{M}, \, \pi_1, \, \iota)$ and
$(\mathcal{B}_2, \, \mathcal{M}, \, \pi_2, \, \jmath)$ be two
geometric fiber bundles over the manifold $\mathcal{M}$.
\begin{defn}
  \label{defn:nth-concoms}
  An \emph{\nth-order concomitant} ($n$ a strictly positive integer)
  from $\mathcal{B}_1$ to $\mathcal{B}_2$ is a smooth mapping
  $\chi : J^n \mathcal{B}_1 \rightarrow \mathcal{B}_2$ such that
  \begin{enumerate}
      \item $( \forall u \in J^n \mathcal{B}_1 )( \forall \phi \in
    \mathfrak{A}_\mathcal{M} ) \; \jmath (\phi) (\chi(u)) = \chi (
    \iota^n (\phi) (u) )$
      \item there is no $(n-1)^{\text{th}}$-order concomitant
    $\chi' : J^{n-1} \mathcal{B}_1 \rightarrow \mathcal{B}_2$
    satisfying
    \[
    ( \forall u \in J^n \mathcal{B}_1 ) \; \chi (u) = \chi'
    (\theta^{n,n-1} (u))
    \]
  \end{enumerate}
\end{defn}
A zeroth-order concomitant (or just `concomitant' for short, when no
confusion will arise), is defined by \ref{def:concom}.  

An important property of concomitants is that, in a limited sense,
they are transitive.
\begin{prop}
  \label{prop:con-composition}  
  If $\chi_1: J^n \mathcal{B}_1 \rightarrow \mathcal{B}_2$ is an
  \nth-order concomitant and
  $\chi_2: \mathcal{B}_2 \rightarrow \mathcal{B}_3$ is a smooth
  mapping, where $\mathcal{B}_1$, $\mathcal{B}_2$ and $\mathcal{B}_3$
  are geometric bundles over the same base space, then
  $\chi_2 \circ \chi_1$ is an \nth-order concomitant if and only if
  $\chi_2$ is a zeroth-order concomitant.
\end{prop}
This follows directly from the fact that inductions are injective
homomorphisms and concomitants respect the fibers.

It will be of physical interest in \S\ref{sec:conds} to consider the
way that concomitants interact with multiplication by a scalar field.
(Since we consider in this paper only concomitants of linear and
affine objects, multiplication of the object by a scalar field is
always defined.)  Towards that end, let us say that a concomitant is
\emph{homogeneous of weight $w$} if for any constant scalar field
$\xi$
\[
\chi (\iota_1 (\phi) (\xi u_1)) = \xi^w \iota_2 (\phi) (\chi (u_1))
\]

\section{Concomitants of the Metric}
\label{sec:concomitants-metric}
\resetsec

As a specific example that will be of use in what follows, consider
the geometric fiber bundle
$(\mathcal{B}_{\text{\small g}}, \, \mathcal{M}, \, \pi_{\text{\small
    g}}, \, \iota_{\text{\small g}})$,
with $\mathcal{M}$ a 4-dimensional, Hausdorff, paracompact, connected,
smooth manifold (\emph{i}.\emph{e}., a candidate spacetime manifold),
the fibers of $\mathcal{B}_{\text{\small g}}$ diffeomorphic to the
space of Lorentz metrics at each point of $\mathcal{M}$, all of the
same signature $(+, \, -, \, -, \, -)$, and $\iota_{\text{\small g}}$
the induction defined by the natural pull-back.  Since the set of
Lorentz metrics in the tangent plane over a point of a 4-dimensional
manifold, all of the same signature, is a 10-dimensional
manifold,\footnote{In fact, it is diffeomorphic to a connected,
  convex, open subset---an open cone with vertex at the origin---in
  $\mathbb{R}^{10}$, and has the further structure of a Fr\'echet
  manifold \cite{curiel-meas-topo-prob-cosmo}.}  the bundle space
$\mathcal{B}_{\text{\small g}}$ is a 14-dimensional manifold.  A
cross-section of this bundle defines a Lorentz metric field on the
manifold.

The following proposition precisely captures the statement one
sometimes hears that there is no scalar or tensorial quantity one can
construct depending only on the metric and its first-order partial
derivatives at a point of a manifold.
\begin{prop}
  \label{prop:no1metric}  
  There is no first-order concomitant from $\mathcal{B}_{\text{\small
      g}}$ to any tensor bundle over $\mathcal{M}$.
\end{prop}
To prove this, it suffices to remark that, given any spacetime
$(\mathcal{M}, \, g_{ab})$ and any two points $p,p' \in \mathcal{M}$,
there are open neighborhoods $U$ of $p$ and $U'$ of $p'$ and a
diffeomorphism $\phi: \; \mathcal{M} \rightarrow \mathcal{M}$, such
that $\phi(p) = p'$, $\phi^\sharp (g'_{ab}) = g_{ab}$ at $p$, and
$\phi^\sharp(\nabla_a g_{bc}) = \nabla_a g_{bc}$ at $p$, where
$\nabla_a$ is any derivative operator other than the Levi-Civita one
associated with $g_{ab}$, and $\phi^\sharp$ is the map naturally
induced by the pull-back action of $\phi$.

This is not to say, however, that no information of interest is
contained in $J^1 \mathcal{B}_{\text{\small g}}$.  Indeed, two metrics
$g_{ab}$ and $h_{ab}$ are first-order osculant at a point if and only
if they have the same associated covariant derivative operator at that
point.  To see this, first note that, if they osculate to first order
at that point, then $\hat{\nabla}_a (g_{bc} - h_{bc}) = 0$ at that
point for all derivative operators.  Thus, for the derivative operator
$\nabla_a$ associated with, say, $g_{ab}$, $\nabla_a (g_{bc} - h_{bc})
= 0$, but $\nabla_a g_{bc} = 0$, so $\nabla_a h_{bc} = 0$ at that
point as well.  Similarly, if the two metrics are equal and share the
same associated derivative operator $\nabla_a$ at a point, then
$\hat{\nabla}_a (g_{bc} - h_{bc}) = 0$ at that point for all
derivative operators, since their difference will be identically
annihilated by $\nabla_a$, and $g_{ab} = h_{ab}$ at the point by
assumption.  Thus they are first-order osculant at that point and so
in the same 1-jet.  This proves that all and only geometrically
relevant information contained in the 1-jets of Lorentz metrics on
$\mathcal{M}$ is encoded in the fiber bundle over spacetime the values
of the fibers of which are ordered pairs consisting of a metric and
the metric's associated derivative operator at a spacetime point.

The second jet bundle over $\mathcal{B}_{\text{\small g}}$ has a
similarly interesting structure.  Clearly, if two metrics are in the
same 2-jet, then they have the same Riemann tensor at the point
associated with the 2-jet, since the result of doubly applying an
arbitrary derivative operator (not the Levi-Civita one associated with
the metric) to it at the point yields the same tensor.  Assume now
that two metrics are in the same 1-jet and have the same Riemann
tensor at the associated spacetime point.  If it follows that they are
in the same 2-jet, then essentially all and only geometrically
relevant information contained in the 2-jets of Lorentz metrics on
$\mathcal{M}$ is encoded in the fiber bundle over spacetime the points
of the fibers of which are ordered triplets consisting of a metric,
the metric's associated derivative operator and the metric's Riemann
tensor at a spacetime point.  To demonstrate this, it suffices to show
that if two Levi-Civita connections agree on their respective Riemann
tensors at a point, then the two associated derivative operators are
in the same 1-jet of the bundle whose base-space is $\mathcal{M}$ and
whose fibers consist of the affine spaces of derivative operators at
the points of $\mathcal{M}$ (because they will then agree on the
result of application of themselves to their difference tensor, and
thus will be in the 2-jet of the same metric at that point).

Assume that, at a point $p$ of spacetime, $g_{ab} = \tilde{g}_{ab}$,
$\nabla_a = \tilde{\nabla}_a$ (the respective derivative operators),
and $R^a {}_{bcd} = \tilde{R}^a {}_{bcd}$ (the respective Riemann
tensors).  Let $C^a {}_{bc}$ be the symmetric difference-tensor
between $\nabla_a$ and $\tilde{\nabla}_a$, which is itself 0 at $p$ by
assumption.  Then by definition $\nabla_{[b} \nabla_{c]} \xi^a = R^a
{}_{bcn} \xi^n$ for any vector $\xi^a$, and so at $p$
\begin{equation*}
  \begin{split}
    R^c {}_{abn} \xi^n &= \nabla_{[a} \tilde{\nabla}_{b]} \xi^c \\
    &= \nabla_a (\nabla_b \xi^c + C^c {}_{bn} \xi^n) -
    \tilde{\nabla}_b  \nabla_a\xi^c \\
    &= \nabla_a \nabla_b \xi^c + \nabla_a (C^c {}_{bn} \xi^n) -
    \nabla_b \nabla_a \xi^c - C^c {}_{bn} \nabla_a \xi^n + C^n {}_{ba}
    \nabla_n\xi^c
  \end{split}
\end{equation*}
but $\nabla_b \nabla_c \xi^a - \nabla_c \nabla_b \xi^a = 2 R^a
{}_{bcn} \xi^n$ and $C^a {}_{bc} = 0$, so expanding the only remaining
term gives
\[
\xi^n \nabla_a C^c {}_{bn} = 0
\]
for arbitrary $\xi^a$ and thus $\nabla_a C^b {}_{cd} = 0$ at $p$; by
the analogous computation, $\tilde{\nabla}_a C^b {}_{cd} = 0$ as well.
It follows immediately that $\nabla_a$ and $\tilde{\nabla}_a$ are in
the same 1-jet over $p$ of the affine bundle of derivative operators
over $\mathcal{M}$.  We have proven
\begin{theorem}
  \label{thm:1-2-jet-metric}
  $J^1 \mathcal{B}_{\text{\small g}}$ is naturally diffeomorphic to
  the geometric fiber bundle over $\mathcal{M}$ whose fibers consist
  of pairs $(g_{ab}, \, \nabla_a)$, where $g_{ab}$ is the value of a
  Lorentz metric field at a point of $\mathcal{M}$, and $\nabla_a$ is
  the value of the covariant derivative operator associated with
  $g_{ab}$ at that point, the induction being defined by the natural
  pull-back.  $J^2 \mathcal{B}_{\text{\small g}}$ is naturally
  diffeomorphic to the geometric fiber bundle over $\mathcal{M}$ whose
  fibers consist of triplets $(g_{ab}, \, \nabla_a, \, R_{abc} {}^d)$,
  where $g_{ab}$ is the value of a Lorentz metric field at a point of
  $\mathcal{M}$, and $\nabla_a$ and $R_{abc}{}^d$ are respectively the
  covariant derivative operator and the Riemann tensor associated with
  $g_{ab}$ at that point, the induction being defined by the natural
  pull-back.
\end{theorem}
It follows immediately that there is a first-order concomitant from
$\mathcal{B}_{\text{\small g}}$ to the geometric bundle
$(\mathcal{B}_\nabla, \, \mathcal{M}, \, \pi_\nabla,$ $\iota_\nabla)$
of derivative operators, \emph{viz}., the mapping that takes each
Lorentz metric to its associated derivative operator.  (This does not
contradict proposition~\ref{prop:no1metric}, as the bundle of
derivative operators is an affine not a tensor bundle.)  Likewise,
there is a second-order concomitant from
$\mathcal{B}_{\text{\small g}}$ to the geometric bundle
$(\mathcal{B}_{\text{\small Riem}}, \, \mathcal{M}, \,
\pi_{\text{\small Riem}},$
$\iota_{\text{\small Riem}})$ of tensors with the same index structure
and symmetries as the Riemann tensor, \emph{viz}., the mapping that
takes each Lorentz metric to its associated Riemann tensor.  (This is
the precise sense in which the Riemann tensor associated with a given
Lorentz metric is ``a function of the metric and its partial
derivatives up to second order''.)  It is easy to see, moreover, that
both concomitants are homogeneous of degree 0.

It follows from theorem~\ref{thm:1-2-jet-metric} and
proposition~\ref{prop:con-composition} that a concomitant of the
metric will be second order if and only if it is a zeroth-order
concomitant of the Riemann tensor:
\begin{prop}
  \label{prop:0-concoms-riemann}
  A concomitant of the metric is second-order if and only if it can be
  expressed as a sum of terms consisting of constants multiplied by
  the Riemann tensor, the Ricci tensor, the Ricci scalar curvature,
  and contractions and products of these with the metric itself.
\end{prop}

\section{Conditions on a Possible Gravitational Stress-Energy Tensor}
\label{sec:conds}
\resetsec

We are almost in a position to state and prove the main result of the
paper, the nonexistence of a gravitational stress-energy tensor.  In
order to formulate and prove a result having that proposition as its
natural interpretation, one must first lay down some natural
conditions on the proposed object, to show that no such object exists
satisfying the conditions.  In general relativity, the stress-energy
tensor is the fundamental invariant quantity encoding all known
localized energetic properties of all known types of matter field, in
the sense that each known type of matter field has a canonical, unique
form of stress-energy tensor associated with it, and all other
invariant energetic quantities associated with the matter field are
derivable from that object.  The canonical form of a stress-energy
tensor is a two-index, symmetric, covariantly divergence-free
tensor.\footnote{\label{fn:other-energy-qs}Thus, the Bel-Robinson
  tensor is ruled out from the start, as it is a 4-index tensor.  (For
  characterization and discussion of the Bel-Robinson tensor and its
  properties, including the way it gives rise to energy-like
  quantities, see
  Senovilla~\citeyearNP{senovilla-superenrgy-tens,senovilla-superergy-tens-apps},
  \citeNP{garecki-rmrks-bel-rob-tens} and
  \citeNP{gomez-lobo-dynal-laws-supenrgy-gr}.)  There are indeed
  several other ``energetic quantities'' that have in general
  relativity invariant representation in some form other than a
  stress-energy tensor, \emph{e}.\emph{g}., the ADM mass and various
  so-called quasi-local quantities
  \cite{szabados-qloc-en-mom-ang-mom-gr}.  Since none of those are
  localized quantities, I do not consider them to be relevant to the
  purposes of this paper.  (One might also reasonably complain, so far
  as my purposes go, that all of those quantities do not differentiate
  between gravitational and non-gravitational forms of energy, but
  rather represent only total, aggregate energy.)  Starting with
  \citeN{komar-cov-cons-laws-gr} and
  \citeN{finkelstein-misner-new-cons-laws}, there is another tradition
  in the context of general relativity of searching for quantities
  that one might hope to be able to interpret as energetic quantities,
  possibly associated in a physically relevant way with the
  ``gravitational field'', \emph{viz}., the search for scalar and
  1-index objects satisfying various forms of ``conservation laws''.
  (See as well, \emph{e}.\emph{g}., \citeNP{trautman-cons-laws-gr} and
  \citeNP{goldberg-inv-trans-cons-laws-ener-mom}.)  As interesting as
  that work is from a mathematical point of view, and as potentially
  interesting as it may be from a physical point of view, I do not
  consider here any of those quantities as viable candidates for
  representations of a localized gravitational energetic quantity, for
  several reasons.  If there are localized energy-like quantities
  associated with ``the gravitational field'' in general relativity
  that do not have the structure of $(0,2)$-index tensor, quantities
  which are found from investigation of various possible forms of
  conservation laws, then it seems to me there are two possibilities:
  there is in fact a gravitational stress-energy tensor, and one can
  derive those quantities from it, even though that is not how they
  were discovered; or those quantities are in fact representative of
  localized gravitational stress-energy, but the claim that they are
  energetic in some important physical sense has to be articulated and
  justified, with a particular eye to explaining how such an
  energy-like quantity interacts with (or not) and is fungible with
  (or not) the stress-energy content of ordinary matter.  I do not
  know how to do it for any of the objects associated with the search
  for single-index conservation laws.  Indeed, it is striking that
  none of the researchers who have investigated such objects discuss
  in any detail the possible physical interpretation of the
  mathematical structures they were investigating, and in particular
  how such quantities may relate to what we understand about ordinary
  stress-energy.}  Not just any such tensor will do, however, for that
gives only the baldest of formal characterizations of it.  From a
physical point of view, at a minimum the object must have the physical
dimension of stress-energy for it to count as a stress-energy tensor.
That it have the dimension of stress-energy is what allows one to add
two of them together in a physically meaningful way to derive the
physical sum of total stress-energy from two different sources.  In
classical mechanics, for instance, both velocity and spatial position
have the form of a three-dimensional vector, and so their formal sum
is well defined, but it makes no physical sense to add a velocity to a
position because the one has dimension \texttt{length/time} and the
other the dimension \texttt{length}.  (I will give a precise
characterization of ``physical dimension'' below.)

An essential, defining characteristic of energy in classical physics
is its obeying some formulation of the First Law of Thermodynamics.
The formulation of the First Law I rely on is somewhat unorthodox:
that all forms of stress-energy are in principle ultimately
fungible---any form of energy can in principle be transformed into any
other form\footnote{\citeN[ch.~\textsc{v}, \S97]{maxwell-matt-mot}
  makes this point especially clearly, including its relation to the
  principle of energy conservation.  See also \citeN[chs.~\textsc{i,
    iii, iv, viii, xii}]{maxwell-theory-heat-1888}.}---not necessarily
that there is some absolute measure of the total energy contained in a
system or set of systems that is constant over time.  In more precise
terms, this means that all forms of energy must be represented by
mathematical structures that allow one to define appropriate
operations of addition and subtraction among them, which the canonical
form of the stress-energy does allow for.\footnote{This kind of linear
  structure is a requirement even if one takes a more traditional view
  of the First Law as making a statement about conservation of a
  magnitude measuring a physical quantity.}  I prefer this formulation
of the First Law in general relativity because there will not be in a
generic spacetime any well-defined global energetic quantity that one
can try to formulate a conservation principle for.  In so far as one
wants to hold on to some principle like the classical First Law in a
relativistic context, therefore, I see no other way of doing it
besides formulating it in terms of fungibility.  (If one likes, one
can take the fungibility condition as a necessary criterion for any
more traditional conservation law.)  This idea is what the demand that
\emph{all} stress-energy tensors, no matter the source, have the same
physical dimension is intended to
capture.\footnote{\label{fn:einstein-seten}For what it's worth, this
  conception has strong historical warrant---Einstein (implicitly)
  used a very similar idea in one of his first papers laying out and
  justifying the general theory \cite[p.~149]{einstein-fndn-gtr}:
  \begin{quote}
    It must be admitted that this introduction of the energy-tensor of
    matter is not justified by the relativity postulate alone.  For
    this reason we have here deduced it from the requirement that the
    energy of the gravitational field shall act gravitatively in the
    same way as any other kind of energy.
  \end{quote}
  \citeN{moller-energy-mom-gr} also stresses the fact that the
  formulation of integral conservation laws in general relativity
  based on pseudo-tensorial quantities depends crucially on the
  assumption that gravitational energy, such as it is, shares as many
  properties as possible with the energy of ponderable
  (\emph{i}.\emph{e}., non-gravitational) matter.}

To sum up, the stress-energy tensor encodes in general relativity all
there is to know of ponderable (\emph{i}.\emph{e}., non-gravitational)
energetic phenomena at a spacetime point:
\begin{enumerate}  
    \item it has 10 components representing with respect to a fixed
  pseudo-orthonormal frame, say, the 6 components of the classical
  stress tensor, the 3 components of linear momentum and the scalar
  energy density of the ponderable field at that point; 
    \item that it has two covariant indices represents the fact that
  it defines a linear mapping from timelike vectors at the point
  (``worldline of an observer'') to covectors at that point
  (``4-momentum covector of the field as measured by that observer''),
  and so defines a bi-linear mapping from pairs of timelike vectors to
  a scalar density at that point (``scalar energy density of the field
  as measured by that observer''), because energetic phenomena,
  crudely speaking, are marked by the fact that they are quadratic in
  velocity and momental phenomena linear in velocity; 
    \item that it is symmetric represents, ``in the limit of the
  infinitesimal'', the classical principle of the conservation of
  angular momentum; it also encodes part of the relativistic
  equivalence of momentum density and the flux of scalar energy
  density;
    \item that it is covariantly divergence-free represents the fact
  that, ``in the limit of the infinitesimal'', the classical
  principles of energy and linear momentum conservation are obeyed; it
  also encodes part of the relativistic equivalence of
  momentum-density and scalar energy density flux; 
    \item the localization of ponderable stress-energy and its
  invariance as a physical quantity are embodied in the fact that the
  object representing it is a \emph{tensor}, a multi-linear map acting
  only on the tangent and cotangent planes of the point it is
  associated with;\footnote{More generally, the notion of localized
    quantity I use here means to be represented by a tensor-like
    object (scalar, tensor, spinor, affine, conformal, projective,
    \ldots), one that has values attributable to individual spacetime
    points and that in some sense or other has properties or actions
    that ramify into the tangent plane over that point in a way that
    can be made sense of by restricting attention to the tangent
    plane.}
    \item finally, the thermodynamic fungibility of energetic
  phenomena is represented by the fact that the set of stress-energy
  tensors forms a vector space---the sum and difference of any two is
  itself a possible stress-energy tensor, and there is a distinguished
  zero element---all elements of which have the same physical
  dimension.
\end{enumerate}
Consequently, the appropriate mathematical representation of localized
gravitational stress-energy, if there is such a thing, is a two
covariant-index, symmetric, covariantly divergence-free tensor having
the physical dimension of
stress-energy.\footnote{\label{fn:pitts}\citeN{pitts-gauge-inv-local-grav-enrgs}
  has proposed an infinite number of ways to define quantities that he
  calls representations of localized gravitational energies (all
  inequivalent).  I exclude Pitts's proposal because I cannot see any
  physical content to his constructed quantities.  How,
  \emph{e}.\emph{g}., could one use one of them to compute the energy
  a gravitational-wave sensor would absorb from ambient gravitational
  radiation?  Precisely because his quantities depend on the frame one
  fixes to make the computation, there can be no invariant, physically
  well defined answer to such a question.  If I stick a rod of
  piezoelectric material in my cup of coffee and use it to warm the
  coffee from the heat it generates by being deformed by a passing
  gravitational wave, then surely the rise in temperature of the
  coffee does not depend on which frame I use to perform the
  calculation.  How should the piezoelectric ``know'' which of Pitts's
  ``localized energies'' it should draw on?  Since there seems to be
  no way to answer such basic physical questions in an unambiguous
  way, I do not see that what he has done is to characterize a
  \emph{physical} quantity.}  (That we demand it be covariantly
divergence-free is a delicate matter requiring further discussion,
which I give at the end of this section.)

Now, in order to make precise the idea of having the physical
dimension of stress-energy, recall that in general relativity all the
fundamental units one uses to define stress-energy, namely time,
length and mass, can themselves be defined using only the unit of time
(or equivalently, using only units of length or mass); these are
so-called geometrized units
\cite[p.~36]{misner-et-grav}.\footnote{\citeN{aldersley-dim-anal-rel-grav}
  contains an interesting discussion of geometrized units, and proves
  a result superficially similar to
  theorem~\ref{thm:only-einstein-tens}, albeit in a very different way
  than I give here.  I have trouble understanding many of his
  arguments and conclusions, however, as he seems to imply that the
  physical dimensions of the components of a quantity depend on the
  physical dimensions of the coordinates in a coordinate system in
  which the quantity is represented.  This makes no sense to me.  A
  quantity simply has a physical dimension, and how one represents it
  in a coordinate system, if one does at all, is physically irrelevant
  to that fact.}  For time, this is trivially true: stipulate, say,
that a time-unit is the time it takes a certain kind of atom to
vibrate a certain number of times under certain conditions.  A unit of
length is then defined as that in which light travels \emph{in vacuo}
one time-unit.  A unit of mass is defined as that of which two, placed
one length-unit apart, will induce in each other by dint of their
mutual gravitation alone an acceleration towards each other of one
length-unit per time-unit per time-unit.\footnote{This definition may
  appear circular, in that it would seem to require a unit of mass in
  the first place before one could say that bodies were of the
  \emph{same} mass.  I think the circularity can be mitigated by using
  two bodies for which there are strong prior grounds for positing
  that they are of equal mass, \emph{e}.\emph{g}., two fundamental
  particles of the same type.  It also suffers from a fundamental lack
  of rigor that the definition of length does not suffer from.  In
  order to make the definition rigorous, one would have to show that
  there exists a solution of the Einstein field-equation
  (approximately) representing two particles in otherwise empty space
  (as defined by the form of $T_{ab}$)---\emph{viz}., two timelike
  geodesics---such that, if on a spacelike hypersurface at which they
  both intersect 1 unit of length apart (as defined on the
  hypersurface with respect to either) they accelerate towards each
  other (as defined by relative acceleration of the geodesics) one
  unit length per unit time squared, then the product of the masses of
  the particles is 1.  I will just assume, for the purposes of this
  paper, that such solutions exist.  Another possibility for
  geometrizing a unit of mass would be to define one as that of a
  Schwarzschild black hole with spatial radius one unit of length, as
  measured with respect to a fixed radial coordinate respecting the
  spherical and timelike symmetries of the spacetime.  It would be of
  some interest to determine the relation between these two different
  ways of defining a geometrized unit of mass.}  These definitions of
the units of mass and length guarantee that they scale in precisely
the same manner as the time-unit when new units of time are chosen by
multiplying the time-unit by some fixed real number
$\lambda^{-\frac{1}{2}}$.  (The reason for the inverse square-root
will become clear in a moment).  Thus, a duration of $t$ time-units
would become $t\lambda^{-\frac{1}{2}}$ of the new units; an interval
of $d$ units of length would likewise become $d\lambda^{-\frac{1}{2}}$
in the new units, and $m$ units of mass would become
$m\lambda^{-\frac{1}{2}}$ of the new units.  This justifies treating
all three of these units as ``the same'', and so expressing
acceleration, say, in inverse time-units.  To multiply the length of
all timelike vectors representing an interval of time by
$\lambda^{-\frac{1}{2}}$, however, is equivalent to multiplying the
metric by $\lambda$ (and so the inverse metric by $\lambda^{-1}$), and
indeed such a multiplication is the standard way one represents a
change of units in general relativity.  This makes physical sense as
the way to capture the idea of physical dimension: all physical units,
the ones composing the dimension of any physical quantity, are
geometrized in general relativity in the most natural formulation, and
so depend only on the scale of the metric itself.  By Weyl's theorem,
however, a metric times a constant represents exactly the same
physical phenomena as the original metric \cite[ch.~2,
\S1]{malament-fnds-gr-ngt}.\footnote{Recall that Weyl's Theorem states
  that the projective structure and the conformal structure determine
  the metric up to a constant.}

Now, the proper dimension of a stress-energy tensor can be determined
by the demand that the Einstein field-equation,
$G_{ab} = \gamma T_{ab}$, where $\gamma$ is Newton's gravitational
constant, remain satisfied when one rescales the metric by a constant
factor.  $\gamma$ has dimension
$\displaystyle \frac{\mbox{(\texttt{length})}^3}
{\mbox{(\texttt{mass}) (\texttt{time})}^2}$, and so in geometrized
units does not change under a constant rescaling of the metric.  Thus
$T_{ab}$ ought to transform exactly as $G_{ab}$ under a constant
rescaling of the metric.  A simple calculation shows that $G_{ab}$
$(= R_{ab} - \half R g_{ab})$ remains unchanged under such a
rescaling.  Thus, a necessary condition for a tensor to represent
stress-energy is that it remain unchanged under a constant rescaling
of the metric.  It follows that the concomitant at issue must be
homogeneous of weight 0 in the metric, whatever order it may be.

We must still determine the order of the required concomitant, for it
is not \emph{a priori} obvious.  In fact, the weight of a homogeneous
concomitant of the metric suffices to fix the differential order of
that concomitant.\footnote{I thank Robert Geroch for pointing this out
  to me.}  This can be seen as follows, as exemplified by the case of
a two covariant-index, homogeneous concomitant $S_{ab}$ of the metric.
A simple calculation based on definition~\ref{defn:nth-concoms} and on
the fact that the concomitant must be homogeneous shows that the value
at a point $p \in \mathcal{M}$ of an \nth-order concomitant $S_{ab}$
can be written in the general form
\begin{equation}
  \label{eq:Sab-form}
  S_{ab} = \sum_\alpha k_\alpha \, g^{qx} \ldots g^{xr} \left(
    \widetilde{\nabla}_x^{(n_1)} g_{qx} \right) \ldots \left(
    \widetilde{\nabla}_x^{(n_i)} g_{xr} \right)
\end{equation}
where: $\widetilde{\nabla}_a$ is any derivative operator at $p$
\emph{other} than the one naturally associated with $g_{ab}$; `$x$' is
a dummy abstract index; `$\widetilde{\nabla}_x^{(n_i)}$' stands for
$n_i$ iterations of that derivative operator (obviously each with a
different abstract index); $\alpha$ takes its values in the set of all
permutations of all sets of positive integers $\{ n_1, \ldots, n_i \}$
that sum to $n$, so $i$ can range in value from 1 to $n$; the
exponents of the derivative operators in each summand themselves take
their values from $\alpha$, \emph{i}.\emph{e}., they are such that
$n_1 + \cdots + n_i = n$ (which makes it an \nth-order concomitant);
for each $\alpha$, $k_\alpha$ is a constant; and there are just enough
of the inverse metrics in each summand to contract all the covariant
indices but $a$ and $b$.

Now, a combinatorial calculation shows
\begin{prop}
  \label{prop:nth-concom-factor}
  If, for $n \geq 2$, $S_{ab}$ is an \nth-order homogeneous
  concomitant of $g_{ab}$, then to rescale the metric by the constant
  real number $\lambda$ multiplies $S_{ab}$ by $\lambda^{n - 2}$.
\end{prop}
In other words, the only such homogeneous \nth-order concomitants must
be of weight $n - 2$.\footnote{The exponent $n - 2$ in this result
  depends crucially on the fact that $S_{ab}$ has only two indices,
  both covariant.  One can generalize the result for tensor
  concomitants of the metric of any index structure.  A slight
  variation of the argument, moreover, shows that there does not in
  general exist a homogeneous concomitant of a given differential
  order from a tensor of a given index structure to one of another
  structure---one may not be able to get the number and type of the
  indices right by contraction and tensor multiplication alone.}  So
if one knew that $S_{ab}$ were multiplied by, say, $\lambda^4$ when
the metric was rescaled by $\lambda$, one would know that it had to be
a sixth-order concomitant.  In particular, $S_{ab}$ does not rescale
when $g_{ab} \rightarrow \lambda g_{ab}$ only if it is a second-order
homogeneous concomitant of $g_{ab}$, \emph{i}.\emph{e}., (by
theorem~\ref{thm:1-2-jet-metric} and
proposition~\ref{prop:0-concoms-riemann}) a zeroth-order concomitant
of the Riemann tensor.  There follows from
proposition~\ref{prop:con-composition}
\begin{lemma}
  \label{lem:riem-0th-concom-0-homog}
  A 2-covariant index concomitant of the Riemann tensor is homogeneous
  of weight zero if and only if it is a zeroth-order concomitant.
\end{lemma}
Thus, such a tensor has the physical dimension of stress-energy if and
only if it is a zeroth-order concomitant of the Riemann tensor.  It is
striking how powerful the physically motivated assumption that the
required object have the physical dimensions of stress-energy is: it
guarantees that the required object will be a second-order concomitant
of the metric.

We now address the issue whether it is appropriate to demand of a
potential gravitational stress-energy tensor that it be covariantly
divergence-free.  In general, I think it is not, even though that is
one of the defining characteristics of the stress-energy tensor of
ponderable matter in the ordinary formulation of general
relativity.\footnote{I thank David Malament for helping me get
  straight on this point.  The following argument is in part
  paraphrastically based on a question he posed to me.}  To see this,
let $T_{ab}$ represent the aggregate stress-energy of all ponderable
matter fields.  Let $S_{ab}$ be the gravitational stress-energy
tensor, which we assume for the sake of argument to exist.  Now, we
ask: can the ``gravitational field'' interact with ponderable matter
fields in such a way that stress-energy is exchanged?  If it could,
then, presumably, there could be interaction states characterized (in
part) jointly by these conditions:
\begin{enumerate}
    \item $\nabla^n (T_{na} + S_{na}) = 0$
    \item\label{item:t-not-cons} $\nabla^n T_{na} \ne 0$
    \item $\nabla^n S_{na} \ne 0$
\end{enumerate}
It is true that, as ordinarily conceived,
condition~\ref{item:t-not-cons} is incompatible with general
relativity as standardly understood and formulated.  The existence of
a gravitational stress-energy tensor, however, would necessarily
entail that we modify our understanding and formulation of general
relativity.  That is why this argument is only \emph{ex hypothesi},
and not meant to be one that would make sense in general relativity as
we actually know it.  (One possible way to understand it,
\emph{e}.\emph{g}., would be that the ways we currently use to
calculate the stress-energy tensor of ordinary matter are mistaken,
precisely in so far as they do not take into account possible
interactions with gravitational phenomena.)

The most one can say, therefore, without wading into some murkily deep
and speculative waters about the way that a gravitational
stress-energy tensor (if there were such a thing) might enter into the
righthand side of the Einstein field-equation or that its existence
might modify the ways we calculate stress-energy for ordinary matter,
is that we expect such a thing would have vanishing covariant
divergence when the aggregate stress-energy tensor of ponderable
matter vanishes, \emph{i}.\emph{e}., that gravitational stress-energy
on its own, when not interacting with ponderable matter, would be
conserved in the sense of being covariantly divergence-free.  This
weaker statement will suffice for our purposes, so we can safely avoid
those murky waters.

Finally, it seems reasonable to require one more condition: were there
a gravitational stress-energy tensor, it should not be zero in any
spacetime with non-trivial curvature, for one can always envision the
construction of a device to extract energy in the presence of
curvature by the use of tidal forces and geodesic deviation.  (See,
\emph{e}.\emph{g}., \citeNP{bondi-mccrea-en-trans-ngt} and
\citeNP{bondi-phys-char-grav-wvs}.)

To sum up:
\begin{condition}
  \label{cond:candidates-sab}
  The only viable candidates for a gravitational stress-energy tensor
  are two covariant-index, symmetric, second-order, zero-weight
  homogeneous concomitants of the metric that are not zero when the
  Riemann tensor is not zero and that have vanishing covariant
  divergence when the stress-energy tensor of ponderable matter
  vanishes.
\end{condition}
This discussion, by the way, obviates the criticism of the claim that
gravitational stress-energy ought to depend on the curvature,
\emph{viz}., that this would make gravitational stress-energy depend
on second-order partial derivatives of the field potential whereas all
other known forms of stress-energy depend only on terms quadratic in
the first partial derivatives of the field potential.  It is exactly
second-order, homogeneous concomitants of the metric that possess
terms quadratic in the first partials.  The rule is that the order of
a homogeneous concomitant is the sum of the exponents of the
derivative operators when the concomitant is represented in the form
of equation~\eqref{eq:Sab-form}.

\section{Gravitational Energy, Again, and the Einstein Field Equation}
\label{sec:nonexist}
\resetsec

\skipline[.5]

\begin{quote}
  If we are to surround ourselves with a perceptual world at all, we
  must recognize as substance that which has some element of
  permanence.  We may not be able to explain how the mind recognizes
  as substantial the world-tensor [\emph{i}.\emph{e}., the Einstein
  tensor], but we can see that it could not well recognize anything
  simpler.  There are no doubt minds which have not this
  predisposition to regard as substantial the things which are
  permanent; but we shut them up in lunatic asylums.
  \begin{flushright}    
    Arthur Eddington \\
    \emph{The Mathematical Theory of Relativity}, pp.~120--121
  \end{flushright}
\end{quote}

\skipline

It follows from lemma~\ref{lem:riem-0th-concom-0-homog}, in
conjunction with condition~\ref{cond:candidates-sab}, that any
candidate gravitational stress-energy tensor must be a zeroth-order
concomitant of $\mathcal{B}_{\mbox{\small Riem}}$, the geometric
bundle of Riemann tensors over spacetime.  (One can take this as a
precise statement of the fact that any gravitational stress-energy
tensor ought to ``depend on the curvature'', as I argued in
\S\ref{sec:princ_equiv}.)  It follows from
proposition~\ref{prop:0-concoms-riemann} that the only possibilities
then are linear combinations of the Ricci tensor and the scalar
curvature multiplied by the metric.  The only covariantly
divergence-free, linear combinations of those two quantities, however,
are constant multiples of the Einstein tensor $G_{ab}$.  (To see this,
note that if there were another, say $k_1 R_{ab} + k_2 R g_{ab}$ for
constants $k_1$ and $k_2$, then
$k_1 R_{ab} + k_2 R g_{ab} - 2k_2 G_{ab}$ would also be divergence
free, but that expression is just a constant multiple of the Ricci
tensor, which is not in general divergence free.)  The Einstein
tensor, however, can still be zero even when the Riemann tensor is not
(when, \emph{e}.\emph{g}., there is only Weyl curvature).  This proves
the main result.
\begin{theorem}
  \label{thm:only-einstein-tens}
  The only two covariant-index, divergence-free concomitants of the
  metric that are homogeneous of zero weight are constant multiples of
  the Einstein tensor.
\end{theorem} 
(Note the strength of the result: not only need one not assume that
the concomitant be second-order, but one need not even assume the
tensor to be symmetric; it all automatically follows from the proof
that all such concomitants of the metric are symmetric.)  Because the
Einstein tensor will be zero in a spacetime having a vanishing Ricci
tensor but a non-trivial Weyl tensor, there follows immediately
\begin{corollary}
  \label{cor:non-exist}
  There are no two covariant-index, divergence-free concomitants of
  the metric that are homogeneous of weight zero that do not
  identically vanish when the Riemann tensor is not zero.
\end{corollary}
The corollary does bear the required natural interpretation, for the
Einstein tensor is not an appropriate candidate for the representation
of gravitational stress-energy: it can be zero in spacetimes with
non-zero curvature; such spacetimes, however, can manifest phenomena,
\emph{e}.\emph{g}., pure gravitational radiation in the absence of
ponderable matter, that one naturally wants to say possess
gravitational energy in some (necessarily non-localized) form or
other.\footnote{As an historical aside, it is interesting to note that
  early in the debate on gravitational energy in general relativity
  \citeN{lorentz-zwaartekracht-3} and \citeN{levi-civita-grav-tensor}
  proposed that the Einstein tensor be thought of as the gravitational
  stress-energy tensor.  Einstein criticized the proposal on the
  grounds that this would result in attributing zero total energy to
  any closed system.}  Non-localizability does mean that gravitational
energy in general relativity, such as it is, is ``nowhere in
particular'', but that is no problem.  The same holds for
gravitational energy (such as it is) in Newtonian theory, and it also
holds for heat in thermodynamics, which is not a localizable quantity,
and more generally for work in classical mechanics.  That does not
mean it is ``not in space-time at all'', no more than any other
globally characterized quantity or entity (\emph{e}.\emph{g}., the
Euler characteristic of the spacetime manifold, or the incompleteness
of an incomplete, inextendible curve, \emph{i}.\emph{e}., a
singularity, or even the ADM mass) is not.  The way such quantities
and entities are ``in space-time'' is a delicate and subtle matter
that does call out for investigation and discussion, but this paper is
not the place for that.  (See \citeNP{curiel-exist-st-struct} for
discussion of the question.)

Theorem~\ref{thm:only-einstein-tens} is similar to the classic result
of \citeN{lovelock-4dim-space-efe}, but it is significantly stronger
in two important ways.\footnote{\label{fn:lovelock}Lovelock proved the
  following, using the definition of concomitant due to Schouten, and
  based on earlier work by \citeN{rund-var-probs-comb-tens} and
  \citeN{duplessis-tens-concoms-cons-laws}.
  \begin{theorem}
    \label{thm:nonexist.lovelock}
    Let $(\mathcal{M},\; g_{ab})$ be a spacetime.  In a coordinate
    neighborhood of a point $p\in \mathcal{M}$, let
    $\Theta_{\alpha \beta}$ be the components of a tensor concomitant
    of
    $\{g_{\lambda \mu} ; \; g_{\lambda \mu,\nu} ; \; g_{\lambda \mu ,
      \nu \rho} \}$ such that
    \[
    \nabla^n \Theta_{nb} = 0.
    \] 
    Then
    \[
    \Theta_{ab} = r G_{ab} + q g_{ab},
    \] 
    where $G_{ab}$ is the Einstein tensor and $q$ and $r$ are
    constants.
  \end{theorem} 
} It does not assume that the desired concomitant be second-order; and
it holds in all dimensions, not just four.  Both of those properties
are grounded on the derivation of the differential order of the
desired concomitant of the metric based on analysis of its required
physical dimension, encoded in the requirement that the concomitant of
the metric be homogeneous of weight zero.  The physical interpretation
of this is that the desired tensor have the physical dimensions of
stress-energy, as is the case for the Einstein tensor, and as must be
the case for any tensor that one would equate to a material
stress-energy tensor to formulate a field equation (so long as the
coupling constant is dimensionless, as is the case for Newton's
constant).  This provides a physical interpretation to the conditions
of the theorem that Lovelock's theorem lacks.

The fact, moreover, that the proof relies essentially only on the
structure of the first and second jet bundles of the bundle of metrics
over a manifold, \emph{i}.\emph{e}., on the bundle of Riemann tensors
over a manifold, and how that structure places severe restrictions on
its possible concomitants, illuminates the physical and geometrical
content of the theorem.  Because Lovelock bases his theorem and its
proof on Schouten's definition of a concomitant, with the attendant
complexity and opacity of the conditions one then has to work with (as
I discussed on p.~\pageref{pg:schouten-difficult}, and in particular
in footnote~\ref{fn:hairy}), his proof consists of several pages of
Baroque and unilluminating coordinate-based, brute-force calculation,
which gives no physical or geometrical insight into why the theorem
holds.  The third difference is that the addition of constant
multiples of the metric is not allowed.  I discuss the consequences of
that below.

Before concluding the paper with a discussion of the bearing of the
theorem on the Einstein field equation, it behooves us to examine a
\emph{prima facie} puzzle my arguments have left us with.  I argued in
\S\ref{sec:conds} that the form of the desired object, that it ought
to be a two-index tensor, followed from the idea that all forms of
stress-energy ought to be fungible, and so \emph{a fortiori} one must
be able to add in a physically significant way entities representing
the stress-energy of different kinds of systems.  Now that I have
shown that there is no gravitational stress-energy tensor, one may be
tempted to conclude that gravitational energy, such as it is, is not
fungible with other forms of energy.  That would be disastrous,
because, as I argued in footnote~\ref{fn:pitts}, there are
circumstances whose only reasonable interpretation is that
gravitational energy, such as it is, is in some way or other being
transformed into other, less \emph{recherch\'e} forms of energy.  (For
more rigorous arguments to this effect, again see
\citeNP{bondi-mccrea-en-trans-ngt} and
\citeNP{bondi-phys-char-grav-wvs}.)  I think the resolution is that,
in general relativity, there is no single framework for analyzing and
interpreting all the phenomena one may want to characterize as
involving the coupling of physical systems based on energy transfer.
Energetic concepts that hang together in a unified framework in
classical physics come apart in general relativity.  When one is
dealing with processes mediated by localizable energetic quantities,
the stress-energy tensor should do the job; otherwise, there are a
multitude of different kinds of quantities any one of which may be
physically relevant to the phenomena at issue.  This should not be
surprising.  We already know of cases in which concepts that formed a
unified framework in classical physics come apart in radical ways in
general relativity, such as the different ways one may characterize a
physical system as being in rotation or not
(Malament~\citeyearNP{malament-rot-nogo,malament-rel-orbit-rot}).  In
any event, even in classical physics there are non-localized energetic
quantities, such as heat in thermodynamics and gravitational potential
energy in Newtonian gravitational theory, that one cannot always treat
in a unified framework with all localized forms of energy, and this
fact never gave rise to any ambiguities in calculations or other
problems.

I conclude the paper by noting that
theorem~\ref{thm:only-einstein-tens} has another reasonable
interpretation, that, in a natural sense the Einstein field equation
is the unique field equation for a theory such as general relativity
that unifies spatiotemporal structure with gravitational phenomena by
way of an appropriate relation between spacetime curvature and the
energetic content of ponderable matter.  (In particular, it follows
from the result that a cosmological-constant term in the field
equation \emph{must} be construed as forming part of the total
stress-energy tensor of spacetime.)
\citeN{malament-newt-grav-geom-spc} makes precise the sense in which
geometrized Newtonian gravity is the limiting theory of general
relativity, as ``the speed of light goes to infinity''.  In
geometrized Newtonian gravity, moreover, the Poisson equation is
formally almost equivalent to the Einstein field equation, and indeed
is identical with it in the vacuum case.  \citeN[ch.~2,
\S7]{malament-fnds-gr-ngt} argues persuasively that, on this basis, it
is natural to adopt the Einstein field equation as the appropriate one
when moving from the context of a Newtonian to a relativistic, curved
spacetime, in so far as any theory better in some sense than Newtonian
theory must, at an absolute minimum, have Newtonian theory as its
limit in certain weak-field regimes.

One can read theorem~\ref{thm:only-einstein-tens} as a way to
generalize this argument.  We know from Newtonian gravitational theory
that the intensity of the gravitational field in a spatial region, in
so far as one can make sense of this idea, is directly proportional to
the density of mass in that region.  In geometrized Newtonian gravity,
this idea is made precise in the geometrized form of the Poisson
equation, which equates a generalized mass-like quantity, which has
the form of a stress-energy tensor, to the Ricci curvature of the
ambient spacetime.  In relativity, one knows that mass just is a form
of energy.  In order for a relativistic theory of gravitation to have
Newtonian gravitational theory as its limiting form, therefore, one is
driven to look for the appropriately analogous equation, equating a
term representing the curvature of a Lorentz metric with a
stress-energy tensor.  Once one imposes natural ancillary conditions
on the desired curvature term, such as that it must be a second-order,
homogeneous concomitant of the metric, then, by
theorem~\ref{thm:only-einstein-tens}, the Einstein field equation
falls out as the only possibility.\footnote{One may take this as a
  more precise and rigorous form of the argument
  \citeN[p.~149]{einstein-fndn-gtr} proposed for his introduction of
  the stress-energy tensor in the first place, as I discussed in
  footnote~\ref{fn:einstein-seten}.}

Theorem~\ref{thm:only-einstein-tens} implies that the addition of
constant multiples of the metric to the geometrical lefthand side of
the Einstein field equation is not allowed.  I interpret that to mean
that any cosmological-constant term must be construed as part of the
total stress-energy tensor of spacetime, and so, in particular, the
cosmological constant itself must have the physical dimensions of
\texttt{mass}$^2$, so that its product with the metric will not change
under constant rescaling of the metric.

In higher dimensions, there are other tensors satisfying Lovelock's
original theorem, the so-called Lovelock tensors.  (Those tensors are
not linear in the second-order partial-derivatives of the metric as
the Einstein tensor is.)  Those tensors form the basis of so-called
Lanczos-Lovelock gravity theories in dimensions higher than four
\cite{lovelock-einst-tens-genls,padmanabhan-kothawala-lanczos-lovelock-mods},
being used to formulate field equations including Lovelock tensors
besides the Einstein tensor.  Because
theorem~\ref{thm:only-einstein-tens} holds in all dimensions, not just
in four, it follows that, in dimensions other than four, the Lovelock
tensors are not homogeneous of weight zero, and so do not have the
physical dimension of stress-energy.  Thus, if one wants to construct
a field equation that equates a linear combination of such tensors to
the stress-energy tensor of ordinary matter, as Lanczos-Lovelock
theories of gravity do, then the coupling constants cannot be
dimensionless like Newton's gravitational constant; the physical
dimension of each coupling constant will be determined by the physical
dimension of the Lovelock tensor it multiplies.  These Lovelock
tensors are usually interpreted as generalizing the Einstein field
equation so as to include curvature terms other than the Einstein
tensor that couple with the stress-energy of ponderable matter.  As in
the case of the cosmological constant, however, the fact that these
Lovelock tensors require dimensionful coupling constants to get the
physical dimensions of the terms right strongly suggests that one
ought not interpret them as geometrical terms coupling to ordinary
stress-energy, but rather as exotic forms of stress-energy themselves.
If this is correct, then Lanczos-Lovelock theories are not in fact
generalizations of general relativity, but rather simply the Einstein
field equation with exotic stress-energy added to the righthand side.
This is an issue that deserves further investigation.


The fact that the same theorem has as its natural interpretation the
uniqueness of the Einstein field equation and the non-existence of a
gravitational stress-energy tensor suggests that there may be a tight
relation between the non-localizability of gravitational stress-energy
and the form of the Einstein field equation.  I have a strong
suspicion this is correct, but I have not been able to put my finger
on exactly what that relation may come to.  A hint, perhaps, comes
from the pregnant remark of \citeN{choquet83} to the effect that the
principle of equivalence (on her interpretation of it) demands that
the gravitational field act as its own source, represented
mathematically by the non-linearity of the Einstein field equation.
Choquet-Bruhat's claim, if true, implies that there can be no linear
field equation for gravity satisfying the equivalence principle, which
would to my mind be a startlingly strong implication for the
equivalence principle to have.  And yet my arguments here suggest that
she may, in some sense, be correct.  That is a question, however, for
future work.

I conclude with an intriguing observation.  The derivation of the
Einstein field equation in \citeN{padmanabhan-thermo-aspects-grav},
based on thermodynamical arguments, is really just a special case of
theorem~\ref{thm:only-einstein-tens} in disguise, as the Einstein
tensor is the only appropriate covariantly divergence-free tensor
having the units of stress-energy, as his proof requires.  (The same
holds true for the generalization of Padmanabhan's arguments to
Lanczos-Lovelock gravity in
\citeNP{padmanabhan-kothawala-lanczos-lovelock-mods}.)  Note,
moreover, that Lovelock's original theorem does not suffice for
Padmanabhan's needs, since it is crucial that the desired tensor have
the right physical dimension.


\begin{thebibliography}{}

\bibitem[\protect\citeauthoryear{Acz\'el}{Acz\'el}{1960}]{aczel-komit-diff-komit}
Acz\'el, J. (1959-1960).
\newblock Ein allgemeines {P}rinzip bez\"uglich {K}omitanten,
  {D}ifferentialkomitanten, kovarianten {A}bleitungen und {A}lgebren von
  \"aquivalenten geometrischen {O}bjekten.
\newblock {\em Acta Classica Universitatis Scientiarum Debreceniensis\/}~{\em
  6\/}(2), 5--13.

\bibitem[\protect\citeauthoryear{Aldersley}{Aldersley}{1977}]{aldersley-dim-anal-rel-grav}
Aldersley, S. (1977, January).
\newblock Dimensional analysis in relativistic gravitational theories.
\newblock {\em Physical Review D\/}~{\em 15\/}(2), 370--376.
\newblock \href{http://dx.doi.org/10.1103/PhysRevD.15.370}
  {doi:10.1103/PhysRevD.15.370}.

\bibitem[\protect\citeauthoryear{Anderson}{Anderson}{1962}]{anderson62}
Anderson, J. (1962).
\newblock Absolute change in general relativity.
\newblock In {\em Recent Developments in General Relativity}, pp.\  121--126.
  Oxford: Pergamon Press.
\newblock Volume commemorating the 60th birthday of Leopold Infeld, with no
  editors mentioned by name.

\bibitem[\protect\citeauthoryear{Anderson}{Anderson}{1967}]{anderson-princs-rel}
Anderson, J. (1967).
\newblock {\em Principles of Relativity Physics}.
\newblock New York: Academic Press.

\bibitem[\protect\citeauthoryear{Ashtekar and Penrose}{Ashtekar and
  Penrose}{1990}]{ashtekar-penrose-mass-pos-focus-struc-sl-inf}
Ashtekar, A. and R.~Penrose (1990, October).
\newblock Mass positivity from focussing and the structure of $i^o$.
\newblock {\em Twistor Newsletter\/}~{\em 31}, 1--5.
\newblock Freely available at
  \url{http://people.maths.ox.ac.uk/lmason/Tn/31/TN31-02.pdf}.

\bibitem[\protect\citeauthoryear{Belot}{Belot}{2011}]{belot-bckgrnd-indep}
Belot, G. (2011, October).
\newblock Background-independence.
\newblock {\em General Relativity and Gravitation\/}~{\em 43\/}(10),
  2865--2884.
\newblock \href{http://dx.doi.org/10.1007/s10714-011-1210-x}
  {doi:10.1007/s10714-011-1210-x}. Preprint:
  \href{http://arxiv.org/abs/1106.0920} {arXiv:1106.0920 [gr-qc].}

\bibitem[\protect\citeauthoryear{Bondi}{Bondi}{1962}]{bondi-phys-char-grav-wvs}
Bondi, H. (1962).
\newblock On the physical characteristics of gravitational waves.
\newblock See \citeN{lichnerowicz-tonnelat-theors-rel-grav}, pp.\  129--135.
\newblock Proceedings of a conference held at Royaumont in June, 1959.

\bibitem[\protect\citeauthoryear{Bondi and McCrea}{Bondi and
  McCrea}{1960}]{bondi-mccrea-en-trans-ngt}
Bondi, H. and W.~McCrea (1960, October).
\newblock Energy tranfer by gravitation in {N}ewtonian theory.
\newblock {\em Mathematical Proceedings of the Cambridge Philosophical
  Society\/}~{\em 56\/}(4), 410--413.
\newblock \href{http://dx.doi.org/10.1017/S0305004100034721}
  {doi:10.1017/S0305004100034721}.

\bibitem[\protect\citeauthoryear{Brading}{Brading}{2005}]{brading-energy-cons-gr}
Brading, K. (2005).
\newblock A note on general relativity, energy conservation, and {N}oether's
  theorems.
\newblock In A.~Kox and J.~Eisenstaedt (Eds.), {\em The Universe of General
  Relativity}, Number~11 in Einstein Studies, pp.\  125--135. Boston:
  Birkh\"auser.

\bibitem[\protect\citeauthoryear{Cartan}{Cartan}{1922}]{cartan22}
Cartan, E. (1922).
\newblock \french{Sur les \'Equations de la Gravitation d'{E}instein}.
\newblock {\em \french{Journal Math\'ematiques et Appliq\'ees}\/}~{\em 1},
  141--203.
\newblock 9 s\'erie.

\bibitem[\protect\citeauthoryear{Choquet-Bruhat}{Choquet-Bruhat}{1983}]{choquet83}
Choquet-Bruhat, Y. (1983).
\newblock Two points of view on gravitational energy.
\newblock In N.~Daruelle and T.~Piran (Eds.), {\em Gravitational Radiation},
  pp.\  399--406. Amsterdam: North Holland Press.
\newblock Proceedings of the NATO Advanced Study Institute, Les Houches Summer
  School on Gravitational Radiation; Les Houches (France), 02--21 Jun 1982.

\bibitem[\protect\citeauthoryear{Curiel}{Curiel}{2017}]{curiel-meas-topo-prob-cosmo}
Curiel, E. (2017).
\newblock Measure, topology and probabilistic reasoning in cosmology.
\newblock Preprint: \href{http://arxiv.org/abs/1509.01878} {arXiv:1509.01878
  [gr-qc]}. Most recent draft available at
  $<$\url{http://strangebeautiful.com/papers/curiel-meas-topo-prob-cosmo.pdf}$>$.

\bibitem[\protect\citeauthoryear{Curiel}{Curiel}{2018}]{curiel-exist-st-struct}
Curiel, E. (2018, June).
\newblock On the existence of spacetime structure.
\newblock {\em British Journal for the Philosophy of Science\/}~{\em 69\/}(2),
  447--483.
\newblock First Published online in 2016.
  \href{http://dx.doi.org/10.1093/bjps/axw014} {doi:10.1093/bjps/axw014}.
  Preprint: \href{http://arxiv.org/abs/1503.03413} {arXiv:1503.03413
  [physics.hist-ph]}. A manuscript containing technical appendices working out
  details of some of the constructions and arguments, and containing further
  discussion of the possible observability of different kinds of spacetime
  structure, is available at
  \url{http://strangebeautiful.com/papers/curiel-exist-st-struct-tech-apdx.pdf}.

\bibitem[\protect\citeauthoryear{{}du Plessis}{{}du
  Plessis}{1969}]{duplessis-tens-concoms-cons-laws}
{}du Plessis, J. (1969, September).
\newblock Tensorial concomitants and conservation laws.
\newblock {\em Tensor\/}~{\em 20}, 347--360.

\bibitem[\protect\citeauthoryear{Eddington}{Eddington}{1923}]{eddington-math-theor-rel}
Eddington, A. (1923).
\newblock {\em Mathematical Theory of Relativity\/} (Second ed.).
\newblock Cambridge: Cambridge University Press.

\bibitem[\protect\citeauthoryear{Einstein}{Einstein}{1915}]{einstein-gr}
Einstein, A. (1915).
\newblock On the general theory of relativity.
\newblock In {\em The Collected Papers of {Albert Einstein} (The Berlin Years:
  Writings, 1914--1917)}, Volume~6, pp.\  98--108. Princeton: Princeton
  University Press.
\newblock Published originally as ``Zur allgemeinen Relativit\"atstheorie'',
  \emph{Sitzungsberichte der K\"oniglich Preu{\ss}ische Akademie der
  Wissenschaften (Berlin)}, Gesamtsitzung 4.\@ November 1915:778--785.

\bibitem[\protect\citeauthoryear{Einstein}{Einstein}{1916}]{einstein-fndn-gtr}
Einstein, A. (1916).
\newblock The foundation of the general theory of relativity.
\newblock In {\em The Principle of Relativity}, pp.\  109--164. New York: Dover
  Press.
\newblock Published originally as ``Die Grundlage der allgemeinen
  Relativit\"atstheorie'', \emph{Annalen der Physik} 49(1916, 7):769--822,
  \href{http://dx.doi.org/10.1002/andp.19163540702}
  {doi:10.1002/andp.19163540702}.

\bibitem[\protect\citeauthoryear{Epstein}{Epstein}{1975}]{epstein-natl-tens-riem-mnflds}
Epstein, D. (1975).
\newblock Natural tensors on {R}iemannian manifolds.
\newblock {\em Journal of Differential Geometry\/}~{\em 10\/}(4), 631--645.
\newblock Permanent link: \url{http://projecteuclid.org/euclid.jdg/1214433166}
  (freely downloadable).

\bibitem[\protect\citeauthoryear{Epstein and Thurston}{Epstein and
  Thurston}{1979}]{epstein-thurston-trans-grps-natl-bunds}
Epstein, D. and W.~Thurston (1979, March).
\newblock Transformation groups and natural bundles.
\newblock {\em Proceedings of the London Mathematical Society\/}~{\em
  s3--38\/}(2), 219--236.
\newblock \href{http://dx.doi.org/10.1112/plms/s3-38.2.219}
  {doi:10.1112/plms/s3-38.2.219}.

\bibitem[\protect\citeauthoryear{Fatibene and Francaviglia}{Fatibene and
  Francaviglia}{2003}]{fatibene-francaviglia-nat-gauge-form-cft}
Fatibene, L. and M.~Francaviglia (2003).
\newblock {\em Natural and Gauge Natural Formalism for Classical Field
  Theories: {A} Geometric Perspective Including Spinors and Gauge Theories}.
\newblock Dordrecht: Kluwer Academic Publishers.

\bibitem[\protect\citeauthoryear{Finkelstein and Misner}{Finkelstein and
  Misner}{1959}]{finkelstein-misner-new-cons-laws}
Finkelstein, D. and C.~Misner (1959, March).
\newblock Some new conservation laws.
\newblock {\em Annals of Physics\/}~{\em 6\/}(3), 230--243.
\newblock \href{http://dx.doi.org/10.1016/0003-4916(59)90080-6}
  {doi:10.1016/0003-4916(59)90080-6}.

\bibitem[\protect\citeauthoryear{Friedman}{Friedman}{1983}]{friedman-fnds-st-theors}
Friedman, M. (1983).
\newblock {\em Foundations of Space-Time Theories: Relativistic Physics and
  Philosophy of Science}.
\newblock Princeton: Princeton University Press.

\bibitem[\protect\citeauthoryear{{Garc\'ia-Parrado
  G\'omez-Lobo}}{{Garc\'ia-Parrado
  G\'omez-Lobo}}{2008}]{gomez-lobo-dynal-laws-supenrgy-gr}
{Garc\'ia-Parrado G\'omez-Lobo}, A. (2008).
\newblock Dynamical laws of supergravity in general relativity.
\newblock {\em Classical and Quantum Gravity\/}~{\em 25\/}(1), 015006.
\newblock \href{http://dx.doi.org/10.1088/0264-9381/25/1/015006}
  {doi:10.1088/0264-9381/25/1/015006}. Preprint:
  \href{http://arxiv.org/abs/arXiv:0707.1475} {arXiv:0707.1475 [gr-qc]}.

\bibitem[\protect\citeauthoryear{Garecki}{Garecki}{2001}]{garecki-rmrks-bel-rob-tens}
Garecki, J. (2001).
\newblock Some remarks on the {B}el-{R}obinson tensor.
\newblock {\em Annalen der Physik\/}~{\em 10\/}(11-12), 911--919.
\newblock
  \href{http://dx.doi.org/10.1002/1521-3889(200111)10:11/12<911::AID-ANDP911>3.0.CO;2-M}
  {doi:10.1002/1521-3889(200111)10:11/12$<$911::AID-ANDP911$>$3.0.CO;2-M}.
  Preprint: \href{http://arxiv.org/abs/gr-qc/0003006} {arXiv:gr-qc/0003006}.

\bibitem[\protect\citeauthoryear{Geroch}{Geroch}{1973}]{geroch-energy-extrac}
Geroch, R. (1973, December).
\newblock Energy extraction.
\newblock {\em Annals of the New York Academy of Sciences\/}~{\em 224},
  108--117.
\newblock Proceedings of the Sixth Texas Symposium on Relativistic
  Astrophysics. \href{http://dx.doi.org/10.1111/j.1749-6632.1973.tb41445.x}
  {doi:10.1111/j.1749-6632.1973.tb41445.x}.

\bibitem[\protect\citeauthoryear{Goldberg}{Goldberg}{1980}]{goldberg-inv-trans-cons-laws-ener-mom}
Goldberg, J. (1980).
\newblock Invariant transformations, conservation laws, and energy-momentum.
\newblock In A.~Held (Ed.), {\em General Relativity and Gravitation}, Volume~1,
  pp.\  469--489. New York: Plenum Press.
\newblock 2 Volumes.

\bibitem[\protect\citeauthoryear{Hirsch}{Hirsch}{1976}]{hirsch-diff-top}
Hirsch, M. (1976).
\newblock {\em Differential Topology}.
\newblock Number~33 in Graduate Texts in Mathematics. New York:
  Springer-Verlag.

\bibitem[\protect\citeauthoryear{Kasner}{Kasner}{1921}]{kasner-geom-thms-efe}
Kasner, E. (1921).
\newblock Geometrical theorems on {E}instein's cosmological equations.
\newblock {\em American Journal of Mathematics\/}~{\em 43}, 217--221.
\newblock \href{http://dx.doi.org/10.2307/2370192} {doi:10.2307/2370192}.

\bibitem[\protect\citeauthoryear{Kol\'a\v{r}, Michor, and Slov\'ak}{Kol\'a\v{r}
  et~al.}{1993}]{kolar-et-nat-opns-dg}
Kol\'a\v{r}, I., P.~Michor, and J.~Slov\'ak (1993).
\newblock {\em Natural Operations in Differential Geometry}.
\newblock Berlin: Springer-Verlag.

\bibitem[\protect\citeauthoryear{Komar}{Komar}{1959}]{komar-cov-cons-laws-gr}
Komar, A. (1959, February).
\newblock Covariant conservation laws in general relativity.
\newblock {\em Physical Review\/}~{\em 113\/}(3), 934--936.
\newblock \href{http://dx.doi.org/10.1103/PhysRev.113.934}
  {doi:10.1103/PhysRev.113.934}.

\bibitem[\protect\citeauthoryear{Landau and Lifschitz}{Landau and
  Lifschitz}{1994}]{landau-lifschitz-fields}
Landau, L. and E.~Lifschitz (1994).
\newblock {\em The Classical Theory of Fields\/} (Fourth revised English
  edition ed.), Volume~2 of {\em Course of Theoretical Physics}.
\newblock Amsterdam: Butterworth Heinemann.
\newblock Translated from the Russian by M. Hammermesh.

\bibitem[\protect\citeauthoryear{Levi-Civita}{Levi-Civita}{1917}]{levi-civita-grav-tensor}
Levi-Civita, T. (1917).
\newblock On the analytic expression that must be given to the gravitational
  tensor in {E}instein's theory.
\newblock \href{http://arxiv.org/abs/arXiv:physics/9906004}
  {arXiv:physics/9906004v1}. A 1999 English translation by S. Antoci and A.
  Loinger of the original published in \emph{Atti della Accademia Nazionale dei
  Lincei, Rendiconti Lincei, Scienze Fisiche e Naturali}, 1917, 26(1):381--391
  (Serie 5).

\bibitem[\protect\citeauthoryear{Lichnerowicz and Tonnelat}{Lichnerowicz and
  Tonnelat}{1962}]{lichnerowicz-tonnelat-theors-rel-grav}
Lichnerowicz, A. and A.~Tonnelat (Eds.) (1962).
\newblock {\em Les Th\'eories Relativistes de la Gravitation}, Number~91 in
  Colloques Internationaux, Paris. Centre National de la Recherche
  Scientifique.
\newblock Proceedings of a conference held at Royaumont in June, 1959.

\bibitem[\protect\citeauthoryear{Lorentz}{Lorentz}{1916}]{lorentz-zwaartekracht-3}
Lorentz, H. (1916).
\newblock Over {E}instein's theorie der zwaartekracht (\textsc{iii}).
\newblock {\em Koninklijke Akademie van Wetenschappen te{} Amsterdam. Verslagen
  van de Gewone Vergaderingen der Wisen Natuurkundige Afdeeling\/}~{\em 25},
  468--486.

\bibitem[\protect\citeauthoryear{Lovelock}{Lovelock}{1971}]{lovelock-einst-tens-genls}
Lovelock, D. (1971).
\newblock The {E}instein tensor and its generalizations.
\newblock {\em Journal of Mathematical Physics\/}~{\em 12\/}(3), 498--501.
\newblock \href{http://dx.doi.org/10.1063/1.1665613} {doi:10.1063/1.1665613}.

\bibitem[\protect\citeauthoryear{Lovelock}{Lovelock}{1972}]{lovelock-4dim-space-efe}
Lovelock, D. (1972).
\newblock The four-dimensionality of space and the {E}instein tensor.
\newblock {\em Journal of Mathematical Physics\/}~{\em 13\/}(6), 874--876.
\newblock \href{http://dx.doi.org/10.1063/1.1666069} {doi:10.1063/1.1666069}.

\bibitem[\protect\citeauthoryear{Malament}{Malament}{1986}]{malament-newt-grav-geom-spc}
Malament, D. (1986).
\newblock {N}ewtonian gravity, limits, and the geometry of space.
\newblock In R.~Colodny (Ed.), {\em From Quarks to Quasars: Philosophical
  Problems of Modern Physics}, pp.\  181--201. Pittsburgh: Pittsburgh
  University Press.

\bibitem[\protect\citeauthoryear{Malament}{Malament}{2002}]{malament-rot-nogo}
Malament, D. (2002).
\newblock A no-go theorem about rotation in relativity theory.
\newblock In D.~Malament (Ed.), {\em Reading Natural Philosophy: Essays in the
  History and Philosophy of Science and Mathematics}. Chicago: Open Court
  Press.
\newblock Essays presented to Howard Stein in honor of his 70th birthday,
  delivered at a \emph{Festchrift} in Stein's honor at the University of
  Chicago, May, 1999.

\bibitem[\protect\citeauthoryear{Malament}{Malament}{2003}]{malament-rel-orbit-rot}
Malament, D. (2003).
\newblock On relative orbital rotation in general relativity.
\newblock In A.~Ashtekar (Ed.), {\em Revisiting the Foundations of Relativistic
  Physics: Festschrift for John Stachel}. Dordrecht: Kluwer.

\bibitem[\protect\citeauthoryear{Malament}{Malament}{2012}]{malament-fnds-gr-ngt}
Malament, D. (2012).
\newblock {\em Topics in the Foundations of General Relativity and Newtonian
  Gravitational Theory}.
\newblock Chicago: University of Chicago Press.
\newblock Uncorrected final proofs for the book are available for download at
  \url{http://strangebeautiful.com/other-texts/malament-founds-gr-ngt.pdf}.

\bibitem[\protect\citeauthoryear{Maxwell}{Maxwell}{1877}]{maxwell-matt-mot}
Maxwell, J.~C. (1877).
\newblock {\em Matter and Motion}.
\newblock New York: Dover Publications, Inc.
\newblock Originally published in 1877. This edition is a 1952 unaltered
  republication of the 1920 Larmor edition.

\bibitem[\protect\citeauthoryear{Maxwell}{Maxwell}{1888}]{maxwell-theory-heat-1888}
Maxwell, J.~C. (1888).
\newblock {\em Theory of Heat}.
\newblock Mineola, NY: Dover Publications, Inc.
\newblock The Dover edition of 2001 republishes in unabridged form the ninth
  edition of 1888 published by Longmans, Green and Co., London, and also
  includes the corrections and notes of Lord Rayleigh incorporated into the
  edition of 1891.

\bibitem[\protect\citeauthoryear{Misner, Thorne, and Wheeler}{Misner
  et~al.}{1973}]{misner-et-grav}
Misner, C., K.~Thorne, and J.~Wheeler (1973).
\newblock {\em Gravitation}.
\newblock San Francisco: Freeman Press.

\bibitem[\protect\citeauthoryear{M{\o}ller}{M{\o}ller}{1962}]{moller-energy-mom-gr}
M{\o}ller, C. (1962).
\newblock The energy-momentum complex in general relativity and related
  problems.
\newblock See \citeN{lichnerowicz-tonnelat-theors-rel-grav}, pp.\  15--29.
\newblock Proceedings of a conference held at Royaumont in June, 1959.

\bibitem[\protect\citeauthoryear{M{\o}ller}{M{\o}ller}{1972}]{moller-thry-rel-2nd}
M{\o}ller, C. (1972).
\newblock {\em The Theory of Relativity\/} (Second ed.).
\newblock Oxford: The Clarendon Press.
\newblock First edition published 1952.

\bibitem[\protect\citeauthoryear{Nijenhuis}{Nijenhuis}{1972}]{nijenhuis-natl-bunds}
Nijenhuis, A. (1972).
\newblock Natural bundles and their general properties.
\newblock In S.~Kobayashi, M.~Obata, and T.~Takahashi (Eds.), {\em Differential
  Geometry: {I}n Honor of {K}entaro Yano}, pp.\  317--334. Tokyo: Kinokuniya
  Book-store Co.

\bibitem[\protect\citeauthoryear{Padmanabhan}{Padmanabhan}{2010}]{padmanabhan-thermo-aspects-grav}
Padmanabhan, T. (2010, March).
\newblock Thermodynamical aspects of gravity: {N}ew insights.
\newblock {\em Reports on Progress in Physics\/}~{\em 73\/}(4), 046901.
\newblock \href{http://dx.doi.org/10.1088/0034-4885/73/4/046901}
  {doi:10.1088/0034-4885/73/4/046901}.

\bibitem[\protect\citeauthoryear{Padmanabhan and Kothawala}{Padmanabhan and
  Kothawala}{2013}]{padmanabhan-kothawala-lanczos-lovelock-mods}
Padmanabhan, T. and D.~Kothawala (2013).
\newblock {L}anczos-{L}ovelock models of gravity.
\newblock {\em Physics Reports\/}~{\em 531\/}(3), 115--171.
\newblock \href{http://dx.doi.org/10.1016/j.physrep.2013.05.007}
  {doi:10.1016/j.physrep.2013.05.007}. Preprint:
  \href{http://arxiv.org/abs/1302.2151} {arXiv:1302.2151 [gr-qc]}.

\bibitem[\protect\citeauthoryear{Palais and Terng}{Palais and
  Terng}{1977}]{palais-terng-nat-bunds-fin-ord}
Palais, R. and C.-L. Terng (1977).
\newblock Natural bundles have finite order.
\newblock {\em Topology\/}~{\em 16\/}(3), 271--277.
\newblock \href{http://dx.doi.org/10.1016/0040-9383(77)90008-8}
  {doi:10.1016/0040-9383(77)90008-8}.

\bibitem[\protect\citeauthoryear{Pauli}{Pauli}{1921}]{pauli-thry-rel}
Pauli, W. (1921).
\newblock {\em The Theory of Relativity}.
\newblock New York: Dover Publications, Inc.
\newblock A 1981 reprint of the 1958 edition from Pergamon Press, a translation
  by G.~Field of the original ``Relativit\"atstheorie'', in
  \emph{Encyklop\"adie der matematischen Wissenschaften}, vol.~V19, B. G.
  Teubner, Leipzig, 1921.

\bibitem[\protect\citeauthoryear{Penrose}{Penrose}{1966}]{penrose-gr-enflux-elem-opt}
Penrose, R. (1966).
\newblock General relativistic energy flux and elementary optics.
\newblock In B.~Hoffman (Ed.), {\em Perspectives in Geometry and General
  Relativity}, pp.\  259--274. Bloomington, IN: Indiana University Press.

\bibitem[\protect\citeauthoryear{Penrose and Rindler}{Penrose and
  Rindler}{1984}]{penrose-rindler-spinors-st-1}
Penrose, R. and W.~Rindler (1984).
\newblock {\em Spinors and Spacetime: Two-Spinor Calculus and Relativistic
  Fields}, Volume~1.
\newblock Cambridge: Cambridge University Press.

\bibitem[\protect\citeauthoryear{Pitts}{Pitts}{2010}]{pitts-gauge-inv-local-grav-enrgs}
Pitts, J. (2010, March).
\newblock Gauge-invariant localization of infinitely many gravitational
  energies from all possible auxiliary structures.
\newblock {\em General Relativity and Gravitation\/}~{\em 42\/}(3), 601--622.
\newblock \href{http://dx.doi.org/10.1007/s10714-009-0874-y}
  {doi:10.1007/s10714-009-0874-y}. Preprint:
  \href{http://arxiv.org/abs/0902.1288} {arXiv:0902.1288 [gr-qc]}.

\bibitem[\protect\citeauthoryear{Rund}{Rund}{1966}]{rund-var-probs-comb-tens}
Rund, H. (1966, June).
\newblock Variational problems involving combined tensor fields.
\newblock {\em Abhandlungen aus dem Mathematischen Seminar der Universit\"at
  Hamburg\/}~{\em 29\/}(3--4), 243--262.
\newblock \href{http://dx.doi.org/10.1007/BF03016051} {doi:10.1007/BF03016051}.

\bibitem[\protect\citeauthoryear{Schouten}{Schouten}{1954}]{schouten-ricci-calc}
Schouten, J. (1954).
\newblock {\em Ricci-Calculus: An Introduction to Tensor Analysis and Its
  Geometrical Applications\/} (Second ed.).
\newblock Number \textsc{x} in Die Grundlehren Der Mathematischen
  Wissenschaften in Einzeldarstellung. Berlin: Springer-Verlag.
\newblock First edition published in 1923, in German. The second edition was
  written in English, and contains significant additions to the first edition.

\bibitem[\protect\citeauthoryear{Schr\"odinger}{Schr\"odinger}{1950}]{schrodinger-st-struc}
Schr\"odinger, E. (1950).
\newblock {\em Space-Time Structure}.
\newblock Cambridge Science Classics. Cambridge: Cambridge University Press.
\newblock Reprinted in 1988.

\bibitem[\protect\citeauthoryear{Senovilla}{Senovilla}{2000}]{senovilla-superenrgy-tens}
Senovilla, J. (2000).
\newblock Super-energy tensors.
\newblock {\em Classical and Quantum Gravity\/}~{\em 17\/}(14), 2799--2842.
\newblock \href{http://dx.doi.org/10.1088/0264-9381/17/14/313}
  {doi:10.1088/0264-9381/17/14/313}. Preprint:
  \href{http://arxiv.org/abs/gr-qc/9906087} {arXiv:gr-qc/9906087}.

\bibitem[\protect\citeauthoryear{Senovilla}{Senovilla}{2002}]{senovilla-superergy-tens-apps}
Senovilla, J. (2002).
\newblock Superenergy tensors and their applications.
\newblock Invited lecture presented at the 1st Conference on Lorentzian
  Geometry, ``Benalmadena 2001''. Preprint:
  \href{http://arxiv.org/abs/math-ph/0202029} {arXiv:math-ph/0202029}.

\bibitem[\protect\citeauthoryear{Steenrod}{Steenrod}{1951}]{steenrod-topo-fbs}
Steenrod, N. (1951).
\newblock {\em The Topology of Fibre Bundles}.
\newblock Number~14 in Princeton Mathematical Series. Princeton, NJ: Princeton
  University Press.

\bibitem[\protect\citeauthoryear{Szabados}{Szabados}{2009}]{szabados-qloc-en-mom-ang-mom-gr}
Szabados, L. (2009).
\newblock Quasi-local energy-momentum and angular momentum in general
  relativity.
\newblock {\em Living Reviews in Relativity\/}~{\em 12}, 4.
\newblock \href{http://dx.doi.org/10.12942/lrr-2009-4}
  {doi:10.12942/lrr-2009-4}. URL (accessed online 11 Sep{} 2015):
  \url{http://www.livingreviews.org/lrr-2009-4}. Update of lrr-2004-04.

\bibitem[\protect\citeauthoryear{Trautman}{Trautman}{1962}]{trautman-cons-laws-gr}
Trautman, A. (1962).
\newblock Conservation laws in general relativity.
\newblock In L.~Witten (Ed.), {\em Gravitation: An Introduction to Current
  Research}, pp.\  169--198. New York: Wiley {\&} Sons Press.

\bibitem[\protect\citeauthoryear{Trautman}{Trautman}{1976}]{trautman-energy-grav-cosmo}
Trautman, A. (1976).
\newblock Energy, gravitation and cosmology.
\newblock In {\em Energy and Physics: Proceedings of the 3rd General Conference
  of the European Physical Society}, pp.\  133--141. Petit-Lancy, Switzerland:
  European Physical Society.

\bibitem[\protect\citeauthoryear{Wald}{Wald}{1984}]{wald-gr}
Wald, R. (1984).
\newblock {\em General Relativity}.
\newblock Chicago: University of Chicago Press.

\bibitem[\protect\citeauthoryear{Weyl}{Weyl}{1921}]{weyl-space-time-matter}
Weyl, H. (1921).
\newblock {\em Space-Time-Matter\/} (Fourth ed.).
\newblock New York: Dover Press.
\newblock A 1952 reprint of the 1950 translation by H. Brose of the 1921
  edition. The first edition published 1918.

\end{thebibliography}
\end{document}